\documentclass[times,authoryear]{elsarticle}

\usepackage[margin=1.5cm]{geometry}
\usepackage{framed,multirow}

\usepackage{amsmath,amssymb}
\usepackage{latexsym}
\usepackage{subfigure}
\usepackage{booktabs,multirow}
\usepackage{xspace}

\usepackage[switch]{lineno}

\usepackage{url}
\usepackage{xcolor}
\definecolor{newcolor}{rgb}{.8,.349,.1}

\usepackage[citebordercolor=white]{hyperref}

\newcommand{\bbm}{\begin{bmatrix}}
\newcommand{\ebm}{\end{bmatrix}}
\newcommand{\mbs}[1]{{\boldsymbol{#1}}}
\DeclareMathAlphabet{\mbf}{OT1}{ptm}{b}{n}
\newcommand{\trans}{{\ensuremath{\mathsf{T}}}} 
\newcommand{\f}{\frac}
\newcommand{\vect}{\underrightarrow}
\newcommand{\ura}[1]{{\underrightarrow{{#1}}}}
\newcommand{\beq}{\begin{equation}}
\newcommand{\eeq}{\end{equation}}
\newcommand{\bdis}{\begin{displaymath}}
\newcommand{\edis}{\end{displaymath}}

\newcommand{\dee}{\textrm{d}}
\newcommand{\mbfbar}[1]{{\bar{\mbf{#1}}}}
\newcommand{\mbftilde}[1]{{\tilde{\mbf{#1}}}}
\newcommand{\onehalf}{\mbox{$\textstyle{\frac{1}{2}}$}}
\DeclareMathOperator*{\minimize}{minimize}
\newcommand{\ie}[0]{{i.e.\@}\xspace}
\newcommand{\p}{\partial}
\newcommand{\mbsbar}[1]{{\bar{\boldsymbol{#1}}}}

\journal{Advances in Space Research}

\begin{document}


\begin{frontmatter}

\title{Solar Cruiser Disturbance Torque Estimation and Predictive Momentum Management}

\author[1]{Ping-Yen Shen}
\author[1]{Ryan J. Caverly\corref{cor1}}
\ead{rcaverly@umn.edu}
\cortext[cor1]{Corresponding author: 
  Tel.: +1-612-625-8000;  
  fax: +0-000-000-0000;}

\affiliation[1]{organization={Department of Aerospace Engineering and Mechanics, University of Minnesota, Twin Cities},
                addressline={110 Union St. SE},
                city={Minneapolis, MN},
                postcode={55455},
                country={USA}}



\begin{abstract}
This paper presents a novel disturbance-torque-estimation-augmented model predictive control (MPC) framework to perform momentum management on NASA's Solar Cruiser solar sail mission.
Solar Cruiser represents a critical step in the advancement of large-scale solar sail technology and includes the innovative use of an active mass translator (AMT) and reflectivity control devices (RCDs) as momentum management actuators. The coupled nature of these actuators has proven challenging in the development of a robust momentum management controller. Recent literature has explored the use of MPC for solar sail momentum management with promising results, although exact knowledge of the disturbance torques acting on the solar sail was required.
This paper amends this issue through the use of a Kalman filter to provide real-time estimation of unmodeled disturbance torques. Furthermore, the dynamics model used in this paper incorporates key fidelity enhancements compared to prior work, including Solar Cruiser's four-reaction-wheel assembly and the offset between its center of mass and center of pressure. More realistic operation scenarios involving the tracking of large angle slew maneuvers under attitude-dependent solar radiation force and torque are also performed to further validate the proposed method compared to prior work.
Simulation results demonstrate that the proposed policy successfully manages angular momentum growth under slew maneuvers that exceed the operational envelope of the current state-of-the-art method. 
The inclusion of the disturbance torque estimate is shown to greatly improve the reliability and performance of the proposed MPC approach. This work establishes a new benchmark for Solar Cruiser's momentum management capabilities and paves the way for MPC-based momentum management of other solar sails making use of an AMT and/or RCDs.
\end{abstract}


\end{frontmatter}



\section{Introduction}

Solar sails have the potential to remove the space exploration limits imposed by traditional propellant-based propulsion, thus unlocking a wide range of missions previously unattainable by conventional spacecraft~\citep{macdonald2011solar,berthet2024space,farres2023propellant,miller2022high,FARRES2019L4L5}. Effectively leveraging the propulsion induced by solar radiation pressure (SRP) and unlocking solar sail travel requires both advancements in the design and deployment of large sail structures~\citep{Vatankhahghadim2021-vt,Hibbert2021-xg,Huang2021-ba} and the concurrent development of advanced control technology~\citep{Chen2023-nw,inness2024controls,Inness2023MM}. 
NASA's Solar Cruiser, which features a massive sail membrane area exceeding $1,600$~m$^2$, is designed to pioneer next-generation space exploration capabilities and enable groundbreaking heliophysics observations~\citep{johnson2019solar,JohnsonLes2020SCTM,johnson2022nasa,pezent2021preliminary}. 

Generating the required propulsion from SRP necessitates precise pointing via attitude control. However, the operation of such large, flexible structures introduces significant control challenges~\citep{BONI2023,fu2015attitude,firuzi2018attitude}. Imperfect sail shapes and structural flexibility induce persistent disturbance torques~\citep{gauvain2023solar} that cause angular momentum accumulation within the onboard reaction wheels (RWs).
This necessitates effective momentum management to desaturate the RWs and prevent a loss of attitude control authority.
Conventional momentum management methods, such as thrusters or magnetic torquers, are unsuitable for long-term, interplanetary, deep-space missions because they either require fuel or are limited to operations near Earth's magnetic field. Innovative actuation methodologies have been developed to adapt to solar sail missions~\citep{wie2004solar2,orphee2018solar,lee2025cablessail}.
On Solar Cruiser, momentum management is achieved using two specialized actuators: the active mass translator (AMT) and reflectivity control devices (RCDs)~\citep{Inness2023MM}.

Solar Cruier's AMT functions as an internal mechanism that shifts the spacecraft's center of mass (CM) relative to the sail's center of pressure (CP) in a plane parallel to the sail surface. This controllable motion produces SRP-induced torques to counteract disturbance torques in the pitch and yaw axes (torques within the plane of the sail) and unload RW angular momentum~\citep{orphee2018solar}. 
The RCDs consist of thin-film membrane pairs, positioned near the tip of each sail boom, set at fixed opposite inclination angles~\citep{heaton2023RCD}.
These devices generate a net roll-axis torque (torque normal to the plane of the sail) by selectively varying the reflectivity of the appropriate RCD membranes via applied voltages, resulting from an imbalance in the differential SRP forces.
A key operational constraint of RCDs is their binary actuation, as they function in an on-off manner, capable only of generating either zero torque or a fixed-magnitude torque in the positive or negative roll direction.

The current design of Solar Cruiser employs a decoupled momentum management strategy, where individual-channel threshold-activated proportional-integral-derivative (PID) controllers command the AMT's two axes and a threshold-based strategy governs the RCDs~\citep{Tyler2024}.
While this approach has been shown to manage angular momentum successfully~\citep{Inness2023MM}, its reliance on purely reactive, threshold-based methods, as well as its neglect of coupled interactions between the AMT motion, RCD input, and the resulting effect on motion in all three axes are significant limitations. 
Specifically, the inability of this state-of-the-art method to optimize the AMT and RCD inputs in a coordinated fashion and proactively prevent RW angular momentum saturation limits its performance under larger slew maneuvers.

The increasing computational capability of modern flight hardware has established model predictive control (MPC) as a viable and practical option for spacecraft attitude determination and control~\citep{di2018real,eren2017model,caverly2020electric,HALVERSON20251}.
MPC is a control methodology that solves an online optimization problem to determine the optimal future sequence of control actions, simultaneously enforcing constraints on both state trajectories and actuator limits, with a receding-horizon. The properties of solar sail dynamics are particularly amenable to this strategy. The low magnitude of SRP ensures slew maneuvers are inherently slow and result in smooth system dynamics.
Moreover, MPC has the potential to enforce the hard constraints associated with RW saturation and optimally allocate the limited control authority associated solar sail actuators through state and input constraints. 
This combination of long time scales available for onboard processing and the need to enforce state and input constraints make MPC an ideal choice for the intricate task of solar sail momentum management.

Prior work by~\cite{shen2025} developed MPC-based momentum management strategies tailored for solar sails equipped with an AMT and RCDs.
They developed a dynamics model that captured key features of the AMT movement, including the resultant time-varying changes in the spacecraft's CM and moment of inertia matrix.
The MPC policy developed in the work of~\cite{shen2025} leveraged its optimization capabilities to handle the actuator constraints and requirements, including the enforcement of on-off RCD actuation and AMT motion rate limits, all while incorporating tuning parameters designed to adjust the trade-off between system performance and control effort.
For real-time onboard implementation, the MPC formulation was posed as a quadratic program (QP), which guarantees fast and robust convergence suitable for the short processing cycles required by the flight computer.
The work of~\cite{shenISSS2025} further examined the critical balance between model fidelity and computational cost when implementing the MPC policy developed by~\cite{shen2025} to determine feasibility for real-time implementation. 
However, these MPC implementations both make unrealistic assumptions that the SRP force and torque acting on the solar sail remain constant and that exact knowledge of the disturbance torque acting on the solar sail is available for use within MPC's prediction model.
These are significant assumptions that limit the practical implementation of the MPC policy proposed by~\cite{shen2025}. The SRP forces and torques acting on the solar sail are attitude-dependent and it is virtually impossible to accurately predict disturbance torques from analytical models due to the unpredictable shape deformation of the solar sail and temporal changes in the sail's optical properties~\citep{Wang2025-ql,gauvain2023solar}. The use of an inaccurate torque model within the MPC framework significantly degrades momentum management performance, negating the purported benefits of the MPC momentum management policy. Another practical limitation of the work of~\cite{shen2025,shenISSS2025} is that their implementations assume that the solar sail is equipped with three RWs aligned with the principal axes of its body-fixed frame. Many spacecraft, including Solar Cruiser, have a 4-RW assembly for redundancy and increased performance, which precludes the use of the methods developed by~\cite{shen2025,shenISSS2025}. Furthermore, the implementation in the work of~\cite{shen2025} assumed that the RWs remained in the same plane as the solar sail's CP, which is not representative of Solar Cruiser's geometry. A final limitation to note in the work of~\cite{shen2025} is that it is only capable of regulating the solar sail to a fixed attitude (i.e., an attitude hold). This does not meet the practical needs of a solar sail mission, which may require performing attitude slews.

To overcome the limitations of prior work, this paper presents a novel MPC-based momentum management policy that incorporates disturbance torque estimation, a 4-RW assembly tailored for the Solar Cruiser mission, an attitude-dependent SRP force and torque, and the ability to track attitude slews. A Kalman filter framework is used to estimate the unmodeled disturbance torques and system model errors in real time, thus enhancing the predictive capability of the MPC. Similar Kalman-filtering approaches have been used in the literature to estimate unknown parameters or terms within a system model~\citep{zenere2018coupling,woodbury2010consider,hayes2025atmospheric,ahmed2024tutorial}.  
For example,~\citet{hayes2025atmospheric} used a Kalman filter to estimate the unknown atmospheric density of a satellite during an orbital reentry, while~\citet{ahmed2024tutorial} estimated the unknown wind acting on a small uncrewed air vehicle. 
Solar Cruiser's 4-RW assembly is accounted for within the proposed MPC implementation through the use of the commonly-used pseudo-inverse RW allocation approach~\citep{leve2015spacecraft,alma9974028382701701}. This provides the MPC prediction model with accurate knowledge of the dynamics of each individual RW, allowing for their operation to be constrained within their saturation limits.

This paper presents four key contributions relative to the state-of-the-art in solar sail momentum management, including prior work on MPC-based methods by~\cite{shen2025,shenISSS2025}. The first contribution is a robust MPC momentum management formulation that uses a Kalman filter to estimate unknown disturbance torques acting on the solar sail. To the best of the knowledge of the authors, this is the first realistically-implementable momentum management policy for a solar sail equipped with an AMT and RCD that outperforms the method of~\citet{Tyler2024}. The second contribution is the incorporation of a 4-RW assembly within an MPC-based momentum management policy. To the best of the knowledge of the authors, this is the first MPC-based momentum management policy to consider a realistic 4-RW assembly. The third contribution is an assessment of the proposed momentum management policy in a realistic simulation of Solar Cruiser's dynamics. Specifically, its CM is located a distance from the sail plane, non-ideal and attitude-dependent SRP forces and torques are considered, and the magnitude of the roll torque generated by the RCDs is modeled as attitude-dependent. All of these effects are meaningful when considering Solar Cruiser's dynamics, as they result in substantial coupling between the system's dynamics and actuation, yet they were not considered in the prior work of~\cite{shen2025}.
The fourth contribution is the augmentation of the proposed MPC policy to track large angle slew maneuvers, which is an important operational requirement of a typical solar sail mission. This improvement expands well-beyond the attitude hold capabilities formulated and demonstrated in the prior work of~\cite{shen2025}.

Details of the nonlinear system dynamics of Solar Cruiser, 
the 4-RW control allocation algorithm, and the momentum management actuators are presented in Section~\ref{sec:Dynamics}. This section also provides the linearized dynamics model used in the Kalman filter and MPC frameworks.
The Kalman filter formulation is presented in Section~\ref{sec:KF}, providing details of how the unmeasurable disturbance torques are estimated.
The MPC formulation is presented in Section~\ref{sec:MPC}, detailing the implementation of estimation-prediction framework and the incorporation of the 4-RW assembly into the MPC prediction model. 
Numerical simulation results are presented in Section~\ref{sec:NumSim}, validating the performance of the proposed estimation-augmented MPC with comparisons to Solar Cruiser's state-of-the-art momentum management method~\citep{Tyler2024}. Results in this section are also presented that demonstrate the effect of actuation thresholds within the proposed momentum management policy on actuation efficiency and observability of the roll-axis disturbance torque. A description of the reference slew maneuver used in this work is included in the Appendix.


\section{Attitude Dynamics and Control Actuation} \label{sec:Dynamics}

NASA's Solar Cruiser uses AMT and RCDs as its momentum management actuators~\citep{Inness2023MM}.
The AMT changes the relative alignment of the CM and CP such that the SRP force acting on the CP results in a corresponding torque about the CM. 
Moving the AMT and appropriately placing the CM/CP offset results in a controllable moment that unloads the accumulated RW angular momentum.
However, the moment of inertia matrix changes when the AMT moves and the mass distribution of the sailcraft changes.
The dynamics are thus coupled with the AMT translation.
This section presents the dynamics and control of the sailcraft, starting with important notation and proceeding with its attitude dynamics, the RW attitude control law, details regarding the momentum management control actuation, and the linearized attitude dynamics model used for the Kalman filter and MPC frameworks.


\subsection{Notation}

The identity matrix of dimension $n\times n$ is denoted as $\mbf{1}_{n \times n}$, while an $n \times m$ matrix of zeros is given by $\mbf{0}_{n \times m}$. 
Physical vectors are denoted as $\ura{v}$. Reference frame $\mathcal{F}_a$ is defined by three orthonormal, dextral physical basis vectors $\ura{a}^1$, $\ura{a}^2$, and $\ura{a}^3$. The physical vector $\ura{v}$ resolved in $\mathcal{F}_a$ is denoted as $\mbf{v}_a = \bbm v_{a1} & v_{a2} & v_{a3} \ebm^\trans$. The position of point $q$ relative to point $z$ is given by $\ura{r}^{qz}$, which is expressed as $\mbf{r}_a^{qz}$ when resolved in reference frame $\mathcal{F}_a$. The cross product operator $(\cdot)^\times$ is used to compute the cross product of two vectors resolved in a particular reference frame. For example, $\ura{u} \times \ura{v}$ resolved in $\mathcal{F}_a$ is computed as $\mbf{u}_a^\times \mbf{v}_a$, where
\bdis
\mbf{u}_a^\times = \bbm 0 & -u_{a3} & u_{a2} \\ u_{a3} & 0 & -u_{a1} \\ -u_{a2} & u_{a1} & 0 \ebm,
\edis
and $\mbf{u}_a = \bbm u_{a1} & u_{a2} & u_{a3} \ebm^\trans$.

The direction cosine matrix (DCM) $\mbf{C}_{ba}$ describes the attitude of reference frame $\mathcal{F}_b$ relative to reference frame $\mathcal{F}_a$. While different attitude parameterizations can be used to describe a DCM, Euler-angle sequences are used in this work due to their ease of physical interpretation and the lack of any large-angle maneuvers that would potentially result in a kinematic singularity. The DCM can be used to express a physical vector in different reference frames. For example, $\mbf{v}_{b} = \mbf{C}_{ba} \mbf{v}_a$.

Within the proposed MPC policy, the subscript $j|t_k$ is used to refer to system states or inputs $j$ time steps ahead of the current time step $t_k$.

\subsection{Solar Cruiser Attitude Dynamics}

\begin{figure}[t!] 
\centering
\subfigure[]
{
        \includegraphics[width=0.6\textwidth]{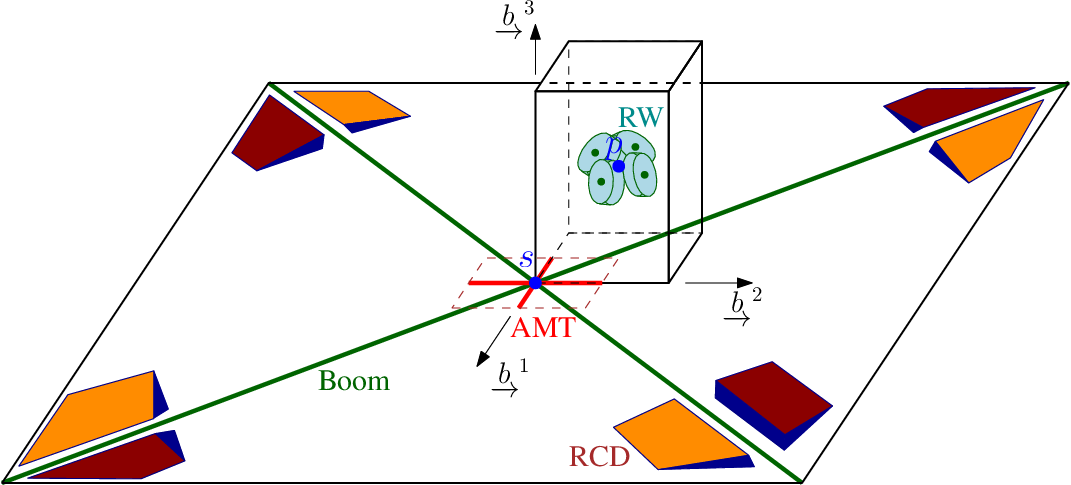}
        \label{Fig:SolarCruiser_diagram}
}
\subfigure[]
{
        \includegraphics[width=0.6\textwidth]{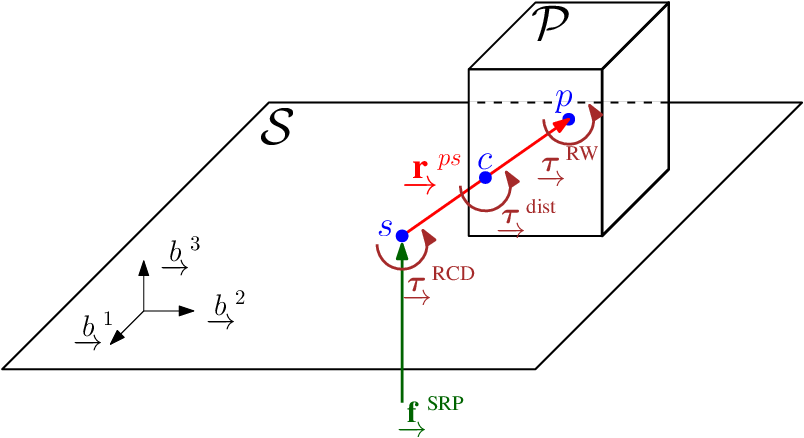}
        \label{Fig:dynamic_model}
}
\centering
\caption{Depictions of the Solar Cruiser model used in this paper (not drawn to scale) highlighting (a) its attitude control and momentum management actuators, including four RWs (light blue), AMT (red) and RCDs (brown and orange); and (b) the definition of key bodies and, such as the sail $\mathcal{S}$ with CM $s$ (also the CP of the entire sailcraft) and the bus $\mathcal{P}$ with CM $s$, as well as the entire sailcraft's CM $c$.}
\label{Fig:dynamic_model_combined}
\end{figure}

Following the approach of~\cite{shen2025} and as illustrated in Fig.~\ref{Fig:dynamic_model_combined}, Solar Cruiser's sail is modeled as a thin flat plate (denoted as $\mathcal{S}$) and a rigid rectangular bus (denoted as $\mathcal{P}$). The nonlinear rigid-body attitude dynamics of a solar sail incorporating the AMT translation as a control input were derived by~\cite{shen2025}, including the time-dependent moment of inertia matrix corresponding to the AMT position.
It is assumed that the each RW spin axis aligns with the position of the RW relative to the CM of the bus (point $p$). Although this assumption is not necessarily required to implement the proposed momentum management method, it removes additional coupling terms in the dynamics that can often be calibrated for upon spacecraft commissioning. As shown in Fig.~\ref{Fig:SolarCruiser_diagram}, the sailcraft body frame $\mathcal{F}_{b}$ is defined with $\vect{b}^3$ pointing through the normal (roll) axis of the sail, and $\vect{b}^2$, $\vect{b}^1$ pointing within the plane of the sail and representing the pitch and yaw axes, respectively. 
The moment of inertia matrix relative to the sailcraft's CM (point $c$) is defined as 
$$
\mbf{J}_b^{\mathcal{B}c}(t) = \mbf{J}^{\mathcal{S}s}_b + \mbf{J}^{\mathcal{P}p}_b - \f{m_p^3+m_s^3}{(m_p+m_s)^2}\mbf{r}_b^{{ps}^\times}(t)\mbf{r}_b^{{ps}^\times}(t),
$$
where $m_s$ and $m_p$ are the masses of the sail and the bus, respectively, $\mbf{J}^{\mathcal{S}s}_b$ and $\mbf{J}^{\mathcal{P}p}_b$ are the nominal moment of inertia matrices of the sail and bus relative to each of their own CMs (points $s$ and $p$), respectively, and the position $\mbf{r}^{ps}_b(t) = \Big[ r^{\text{AMT}}_{b1}(t) \,\,\, r^{\text{AMT}}_{b2}(t) \,\,\, {r}^{ps}_{b3} \Big]^\trans$ contains the controllable AMT positions $r^{\text{AMT}}_{b1}(t)$ and $r^{\text{AMT}}_{b2}(t)$. These AMT positions are actuated via the red linear actuators visualized in Fig.~\ref{Fig:SolarCruiser_diagram}. 
The constant component ${r}^{ps}_{b3}$ represents the constant offset distance in the $\ura{b}^3$ sail-normal direction between the CM of the sail and the CM of the bus. 

Let $\mathcal{F}_{a}$, defined by basis vectors $\vect{a}^1$, $\vect{a}^2$, $\vect{a}^3$, be an inertial reference frame. The solar sail attitude dynamics with mass translation are defined as~\citep{shen2025}
\beq
    \mbf{J}_b^{\mathcal{B}c} \dot{\mbs{\omega}}_b^{ba} + \mbs{\omega}_b^{ba^\times} \mbf{J}_b^{\mathcal{B}c} \mbs{\omega}_b^{ba} + \mbs{\omega}_b^{ba^\times}\mbf{h}_b^{\text{RW}} + \dot{\mbf{h}}_b^{\text{RW}} 
    + \f{m_p^3+m_s^3} {(m_p+m_s)^2} \bigg( \mbf{r}_b^{ps^\times}\ddot{\mbf{r}}_b^{ps} -2\dot{\mbf{r}}_b^{ps^\times} \mbf{r}_b^{ps^\times} \mbs{\omega}_b^{ba} \bigg) = \mbs{\tau}_b^{\mathcal{B}c} , \label{EOM}
\eeq
where 
$\mbs{\omega}_b^{ba}$ is the angular velocity of $\mathcal{F}_{b}$ relative to $\mathcal{F}_{a}$ resolved in $\mathcal{F}_{b}$, 
$\mbf{h}_b^{\text{RW}}$ is the collective angular momentum of the RWs relative to the CM of the bus (point $p$) with respect to inertial reference frame $\mathcal{F}_a$ resolved in $\mathcal{F}_{b}$.
The term $\mbs{\tau}_b^{\mathcal{B}c} = \mbs{\tau}_b^{\text{AMT}} + \mbs{\tau}_b^{\text{RCD}} + \mbs{\tau}_b^{\text{dist}} $ is the torque acting on the sailcraft relative to its CM, consisting of the AMT-SRP-induced torque $\mbs{\tau}_b^{\text{AMT}} = \f{m_s}{m_p+m_s}\mbf{r}_b^{ps^\times} \mbf{f}_b^{\text{SRP}}$, the RCD torque $\mbs{\tau}_b^{\text{RCD}}$, and the disturbance torque $\mbs{\tau}_b^{\text{dist}}$. The term $\mbf{f}_b^{\text{SRP}}$ denotes the solar radiation pressure induced force acting on the sail's CP.
The RCD torque is generated by activating one of the set of four RCDs (either brown or orange) visualized in Fig.~\ref{Fig:SolarCruiser_diagram}. Actuating the orange RCDs generates a positive roll torque about $\vect{b}^3$ axis, while actuating the brown RCDs generates a negative torque about this same axis. 
The time-dependent argument (t) for the variables $\mbf{J}_b^{\mathcal{B}c}(t)$, $\mbf{r}^{ps}_b(t)$, $\mbs{\omega}_b^{ba}(t)$, $\mbf{h}_b^{\text{RW}}(t)$, $\mbs{\tau}_b^{\mathcal{B}c}(t)$ is omitted for brevity in Eq.~\eqref{EOM}.

A 3-2-1 Euler angle sequence is used to describe the rotation between $\mathcal{F}_{a}$ and $\mathcal{F}_{b}$, so that $\mbf{C}_{ba} = \mbf{C}_1(\theta_1) \mbf{C}_2(\theta_2) \mbf{C}_3(\theta_3)$ is the DCM describing the orientation of $\mathcal{F}_{b}$ relative to $\mathcal{F}_{a}$, where $\mbf{C}_i(\cdot)$ is the DCM representing a rotation about the $i$-th principal axis. The mapping $\mbs{\omega}^{ba}_b = \mbf{S}(\mbs{\theta})\dot{\mbs{\theta}} $ relates Euler angle rates to angular velocity, where the column matrix $\mbs{\theta} = \Big[ \theta_1 \,\,\, \theta_2 \,\,\, \theta_3 \Big]^\trans$ is the set of Euler angles and $\mbf{S}(\mbs{\theta})$ is the mapping matrix.
It is worth noting that $\mbf{S}(\mbs{\theta})$ depends only on $\theta_1$ and $\theta_2$ due to the selected 3-2-1 Euler angle sequence. Given that a solar sail is designed to keep the Sun within its field of view and maintain a nominal spin about the $\vect{b}^3$ axis, this choice allows for ease of linearization about any nominal angular velocity about the $\vect{b}^3$ axis, and positions the kinematic singularity at $180\text{\textdegree}$ from the nominal inertial pointing attitude. 

Generally, the forces and torques induced by SRP vary with the attitude of the solar sail, the shape of the sail membrane, and its optical properties. Given the slow nature of solar sail maneuvers, it is assumed that vibrations in the structure are minimal and both the sail shape and material properties remain static, which implies that $\mbf{f}_b^{\text{SRP}}$ and $\mbs{\tau}_b^{\text{dist}}$ depend solely on the attitude of the sailcraft. The optical properties of the Solar Cruiser membrane~\citep{heaton2015update,heaton2017near} and its expected deformations~\citep{gauvain2023solar} are incorporated into the static membrane shape model developed by~\cite{bunker2026static}. This model is then employed to compute $\mbf{f}_b^{\text{SRP}}$ and $\mbs{\tau}_b^{\text{dist}}$ based on the solar sail's attitude $\mbs{\theta}$ in this work.


\subsection{Reaction Wheel Control and Allocation} \label{sec:RW}

The RWs onboard the sailcraft generate the vehicle's attitude control torques through an increase or decrease in the angular momentum of the RWs. Solar Cruiser has a 4-RW assembly, which requires control allocation to determine the action to be taken by each individual RW in order to generate the required attitude control torque. Details regarding the attitude control law, the RW geometric configuration, and the RW control allocation methodology used in this work are presented in this section.

\subsubsection{Attitude Control Law}

Many advanced RW attitude control methods exist that could be implemented to meet the solar sail's attitude control requirements. 
This paper employs a simple PID attitude control law to mimic the controller developed for Solar Cruiser~\citep{Inness2023MM}.
The desired control torque to be generated by the RWs is defined through a PID control law as
\beq
    \mbs{\tau}^{\text{RW}}_{b,\text{des}} = -\dot{\mbf{h}}^\text{RW}_{b,\text{des}} = -\mbf{K}_p \bigg(\mbs{\theta}(t) - \mbs{\theta}_d(t)\bigg) -\mbf{K}_d \bigg(\mbs{\omega}^{ba}_b(t) - \mbs{\omega}_d(t)\bigg) -\mbf{K}_i \int^t_{t_0} \bigg(\mbs{\theta}(\tau) - \mbs{\theta}_d(\tau)\bigg) \dee\tau, \label{eq:PID_law}
\eeq
where $\mbs{\theta}_d(t)$ and $\mbs{\omega}_d(t) = \mbf{S}^{-1}(\mbs{\theta}_d(t))\dot{\mbs{\theta}}_d(t)$ are the desired Euler angles and angular velocity of the desired trajectory, respectively. In this work, $\mbs{\theta}_d$ and $ \dot{\mbs{\theta}}_d$ are chosen to be a smooth trajectory matching a rest-to-rest maneuver with a trapezoidal Euler angle rate profile. Details of the time-dependent desired trajectory can be found in the Appendix. A different trajectory can be selected based on specific mission scenarios.

\subsubsection{Reaction Wheels Assembly Geometry}

Solar Cruiser uses four RWs as its primary attitude control actuators~\citep{Inness2023MM}.
The attitude dynamics, as established in Eq.~\eqref{EOM} within the sailcraft's body frame, include the three-dimensional total angular momentum of the four RWs, $\mbf{h}_b^{\text{RW}}$, as well as its time derivative $ \dot{\mbf{h}}_b^{\text{RW}}$. The variables $\mbf{h}_b^{\text{RW}}$ and $ \dot{\mbf{h}}_b^{\text{RW}}$ represent projections of the angular momentum of the 4-RW configuration onto the three body-frame axes. This results in the linear relationships 
$$
\mbf{h}_b^{\text{RW}} = \mbf{M}_{34}\mbf{h}^{\text{RW}}_4,
$$
and
$$
\dot{\mbf{h}}_b^{\text{RW}} = \mbf{M}_{34}\dot{\mbf{h}}^{\text{RW}}_4,
$$
where $\mbf{h}^{\text{RW}}_4 = \bbm h_1 &h_2&h_3&h_4 \ebm^\trans$ comprises the four individual RW angular momentum values. The allocation matrix $ \mbf{M}_{34} \in \mathbb{R}^{3\times4}$ is time invariant and is determined entirely by the geometric configuration of the four RWs. 
The time-derivative of the total RW angular momentum projected onto the body frame directly yields the reaction torque exerted on the spacecraft, where $\dot{\mbf{h}}^\text{RW}_{b} = -\mbs{\tau}^{\text{RW}}_b $.

The optimal geometric configuration of a 4-RW assembly has been investigated extensively in the literature~\citep{ismail2010study,Bellar2016,lee2017shorter,leve2015spacecraft,alma9974028382701701}. In the absence of any direct information regarding the configuration used by Solar Cruiser, the common pyramidal configuration is used, where it is assumed that the spin axis of each RW passes through the CM of the sailcraft bus. 
This choice of RW configuration only affects the definition of $\mbf{M}_{34}$, allowing the methods presented in this paper to be adapted to other configurations if desired.

To derive the expression for $\mbf{M}_{34}$, consider the $i$-th RW as a rigid disk rotating about its axis of symmetry $\ura{w^3_i}$, in its rotating frame $\mathcal{F}_{w_i}$. The reference frame $\mathcal{F}_{w_i}$ is obtained from $\mathcal{F}_{b}$ by rotating $\psi_i$ about $\ura{b^3}$, then rotating $\phi_i$ about the rotated $\ura{b^2}$ axis. In this paper $\psi_i = 60^\circ$ for all $i$, and $\phi_i = 45^\circ+(i-1)\cdot90^\circ$, $i = 1,2,3,4$.
The DCM defining the orientation of $\mathcal{F}_{w_i}$ relative to the body frame $\mathcal{F}_{b}$ is given by 
\bdis
\mbf{C}_{w_ib} = \mbf{C}_{bw_i}^\trans = \mbf{C}_2(\phi_i) \mbf{C}_3(\psi_i) = \bbm \cos\phi_i\cos\psi_i &\cos\phi_i\sin\psi_i &-\sin\phi_i \\
-\sin\psi_i & \cos\psi_i &0 \\
\sin\phi_i\cos\psi_i &\sin\phi_i\sin\psi_i &\cos\phi_i \ebm. 
\edis
The angular velocity of $\mathcal{F}_{w_i}$ relative to $\mathcal{F}_b$ expressed in $\mathcal{F}_b$ is
\bdis
\mbs{\omega}_b^{w_ib} = \mbf{C}_{bw_i} \bbm 0\\0\\\dot{\gamma}_i \ebm = \bbm \sin{\phi_i}\cos{\psi_i}\\\sin{\phi_i}\sin{\psi_i}\\ \cos{\phi_i}\ebm \dot{\gamma}_i,
\edis
where $\dot{\gamma}_i$ denotes the spin rate of the $i$-th RW.
The moment of inertia matrix of the $i$-th RW is given by $\mbf{J}_{w_i}^{\mathcal{W}_ip} = \text{diag}(\f{m_w r_w^2}{4}, \f{m_w r_w^2}{4}, \f{m_w r_w^2}{2})$, where $m_w$ is the mass of the RW and $r_w$ is the radius of RW. 
The total angular momentum of the four RWs projected into the body frame is obtained by summing the individual contributions, which establishes the final kinematic mapping
\begin{align*}
    \mbf{h}^\text{RW}_b &= \sum^{4}_{i=1} \mbf{C}_{bw_i} \mbf{J}_{w_i}^{\mathcal{W}_ip} \mbf{C}_{bw_i}^\trans \mbs{\omega}_b^{w_ib}  \\
    &= \sum^{4}_{i=1} \mbf{C}_{bw_i} \bbm \f{1}{2} & 0& 0\\ 0 &\f{1}{2} &0 \\ 0 &0 &1 \ebm \mbf{C}_{bw_i}^\trans \bbm \sin{\phi_i}\cos{\psi_i}\\\sin{\phi_i}\sin{\psi_i}\\ \cos{\phi_i}\ebm h_i \\
    &= \mbf{M}_{34}\mbf{h}^{\text{RW}}_4 ,
\end{align*}
where $h_i = \f{1}{2}m_w r_w^2 \dot{\gamma}_i$ is the angular momentum of the $i$-th RW resolved in frame $\mathcal{F}_{w_i}$ and
\bdis
\mbf{M}_{34} = \bbm \mbf{C}_{bw_1} \bbm \f{1}{2} & 0& 0\\ 0 &\f{1}{2} &0 \\ 0 &0 &1 \ebm \mbf{C}_{bw_1}^\trans \bbm \sin{\phi_1}\cos{\psi_1}\\\sin{\phi_1}\sin{\psi_1}\\ \cos{\phi_1}\ebm & \cdots & \mbf{C}_{bw_4} \bbm \f{1}{2} & 0& 0\\ 0 &\f{1}{2} &0 \\ 0 &0 &1 \ebm \mbf{C}_{bw_4}^\trans \bbm \sin{\phi_4}\cos{\psi_4}\\\sin{\phi_4}\sin{\psi_4}\\ \cos{\phi_4}\ebm \ebm.
\edis
Substituting in the numerical parameters $\psi_i = 60^\circ$ for all $i$, and $\phi_i = 45^\circ+(i-1)\cdot90^\circ$, $i = 1,2,3,4$ results in
\beq
\label{eq:RW_mapping_matrix}
\mbf{M}_{34} = \bbm 0.6124&-0.6124&-0.6124&0.6124\\0.6124&0.6124&-0.6124&-0.6124\\0.5&0.5&0.5&0.5\ebm.
\eeq

\subsubsection{Unconstrained Minimum-Norm Allocation}
The PID control law in Eq.~\eqref{eq:PID_law} determines the desired angular momentum derivative in the body-frame $\dot{\mbf{h}}^\text{RW}_{b,\text{des}}$, which then needs to be allocated to the momentum of the indivitual RWs within the 4-RW assembly. Since the mapping between $\mbf{h}_b^{\text{RW}}$ and $\mbf{h}^{\text{RW}}_4$ is under-determined (i.e., four variables are to be determined from three equations), the allocation problem is inherently non-unique. 

The selection of an optimal allocation is typically a core redundancy and safety design choice within the attitude determination and control system (ADCS).
For the purpose of developing momentum management techniques in this paper, a computationally-efficient pseudo-inverse method is employed to define this allocation uniquely, where
\begin{align} 
    \mbf{h}^{\text{RW}}_{4}  &= \mathbf{M}_{34}^\dagger \mbf{h}_{b,\text{des}}^{\text{RW}}, \label{eq:h4_pseu} \\
    \dot{\mbf{h}}^{\text{RW}}_{4} &= \mbf{M}_{34}^\dagger \dot{\mbf{h}}_{b,\text{des}}^{\text{RW}}, \label{eq:h4dot_pseu}
\end{align}
and $\mbf{M}_{34}^\dagger = \mbf{M}_{34}^\trans \Big(\mbf{M}_{34}\mbf{M}_{34}^\trans \Big) ^{-1}$.
This approach yields the minimum-norm pseudo-inverse result for $\mbf{h}^{\text{RW}}_4$ and $\dot{\mbf{h}}^{\text{RW}}_4$, characterized by the smallest possible Euclidean norm ($||\mbf{h}^{\text{RW}}_4||_2$ and $||\dot{\mbf{h}}^{\text{RW}}_4||_2$) amongst all potential combinations that yield the desired values of $\mbf{h}^{\text{RW}}_{b,\text{des}}$ and $\dot{\mbf{h}}^{\text{RW}}_{b,\text{des}}$. In the absence of any RW saturation, Eq.~\eqref{eq:h4dot_pseu} is used to compute $\dot{\mbf{h}}^{\text{RW}}_{4}$.

\subsubsection{Constrained Minimum-Norm Allocation via Sequential Pseudo-Inverse}
The standard pseudo-inverse solution from Eq.~\eqref{eq:h4_pseu} may generate wheel momentum commands that exceed the physical saturation limit of one or more RWs (i.e., $||\mbf{h}^{\text{RW}}_{4}||_\infty > h^{\text{RW}}_{\text{max}}$, where $h^{\text{RW}}_{\text{max}}$ is the saturation limit). 
A constraint-prioritized sequential allocation scheme is applied to manage the inherent redundancy while rigorously enforcing these physical saturation constraints. This scheme is designed to find a solution to the minimization of $||\mbf{h}^{\text{RW}}_4||_2$ subject to the constraint $||\mbf{h}^{\text{RW}}_{4}||_\infty \leq h^{\text{RW}}_{\text{max}}$ in a computationally-efficient manner. Although this could be posed as a QP, the relatively short time steps associated with attitude control necessitates a more computationally-efficient strategy. To meet this need, a suboptimal solution is found through the proposed method that operates by sequentially checking the maximum individual RW angular momentum, enforcing saturation limit, and redistributing angular momentum on unsaturated RWs to satisfy the desired RW attitude control torque.
The process is as follows: 
\begin{enumerate}
    \item \textbf{Unconstrained Initial Solution and Saturation Check:} The process commences by calculating the unconstrained minimum-norm solution using the current body-frame angular momentum $\mbf{h}^{\text{RW}}_{4} = \mbf{M}_{34}^{\dagger} \mbf{h}_{b}^{\text{RW}}$. 
        The resulting unconstrained 4-RW momentum is checked against the saturation limit, where $||\mbf{h}^{\text{RW}}_{4}||_\infty \leq h^{\text{RW}}_{\text{max}}$ must be satisfied.
        If $||\mbf{h}^{\text{RW}}_{4}||_\infty \leq h^{\text{RW}}_{\text{max}}$ is satisfied (none of the RWs saturate), the mapping of the angular momentum derivative is $\dot{\mbf{h}}^{\text{RW}}_{4} = \mbf{M}_{34}^{\dagger} \dot{\mbf{h}}_{b,\text{des}}^{\text{RW}}$, the constrained minimum-norm allocation is determined, and the remaining steps can be skipped.
        If $||\mbf{h}^{\text{RW}}_{4}||_\infty > h^{\text{RW}}_{\text{max}}$ (at least one of the RWs saturate), the process continues to Steps 2 through 4.
    \item \textbf{Saturation Implementation:} The component of $\mbf{h}^{\text{RW}}_{4}$ with the largest magnitude exceeding the saturation limit $h^{\text{RW}}_{\text{max}}$ is identified. This is labeled as the $i$-th component of $\mbf{h}^{\text{RW}}_{4}$ (i.e., $h^{\text{RW},(i)}_{4}$), where $|h^{\text{RW},(i)}_{4}| = ||\mbf{h}^{\text{RW}}_{4}||_\infty$ and $|h^{\text{RW},(i)}_{4}| > h^{\text{RW}}_{\text{max}}$. The $i$-th RW is now identified as saturated for all remaining iterations. 
        Then, its momentum is fixed at the saturation boundary $ h^{\text{RW},(i)}_{4,\text{sat}} = \pm h^{\text{RW}}_{\text{max}}$, where the sign of the momentum $h^{\text{RW},(i)}_{4}$ is maintained. 
        Crucially, when a wheel's angular momentum is saturated and fixed, its corresponding commanded angular momentum derivative $\dot{h}^{\text{RW},(i)}_{4}$ must simultaneously be set to zero to prevent the controller from commanding further change into the limit, \ie, $\dot{h}^{\text{RW},(i)}_{4,\text{sat}} = 0$.
    \item \textbf{Residual Calculation and Redistribution:} The momentum contribution from the saturated wheel(s) is calculated and subtracted from the original demanded body-frame angular momentum. This yields the residual momentum requirement for the remaining unsaturated wheels 
        \bdis
            \mbf{h}_{b}^{\text{res}} = \mbf{h}_{b,\text{des}}^{\text{RW}} - \mbf{M}_{34}^{\text{sat}} \mbf{h}^{\text{RW}}_{4,\text{sat}} ,
        \edis
        where $\mathbf{M}_{34}^{\text{sat}}$ is a modified version of the allocation matrix, where the columns corresponding to the unsaturated wheels are set to zero and $\mbf{h}^{\text{RW}}_{4,\text{sat}} \in \mathbb{R}^4$ contains $\pm h^{\text{RW}}_{\text{max}}$ in the entries associated with saturated wheels and zeros in the other entries of the matrix.
        
        The angular momentum of the $n$ unsaturated wheels is then recalculated as
        \beq \label{eq:RW_redistribution}
            \mbf{h}^{\text{RW}}_{n,\text{unsat}} = \Big(\mbf{M}_{3 \times n}^{\text{unsat}}\Big)^{\dagger} \mbf{h}_{b}^{\text{res}},
        \eeq
        where $\mbf{M}_{3 \times n}^{\text{unsat}}$ is a modified version of the allocation matrix $\mbf{M}_{34}$ such that the columns associated with the saturated wheels are removed, reducing its dimension to $3 \times n$. 
        The allocation of the unsaturated wheels angular momentum rate is computed similarly as
        \beq
        \label{eq:RW_rate_redistribution}
        \dot{\mbf{h}}^{\text{RW}}_{n, \text{unsat}} = \Big(\mbf{M}_{3 \times n}^{\text{unsat}}\Big)^{\dagger} \dot{\mbf{h}}_{b,\text{des}}^{\text{RW}}.
        \eeq
    \item \textbf{Saturation Assessment:} The results of Steps 2 and 3 are compiled to obtain updated values of $\mbf{h}^{\text{RW}}_{4}$ and $\dot{\mbf{h}}^{\text{RW}}_{4}$. The entries of $\mbf{h}^{\text{RW}}_{4}$ associated with saturated wheels are set using the appropriate entries of $\mbf{h}^{\text{RW}}_{4,\text{sat}}$, while the unsaturated wheel values are found using the result from Eq.~\eqref{eq:RW_redistribution}. The entries of $\dot{\mbf{h}}^{\text{RW}}_{4}$ associated with saturated wheels are set to zero, while the unsaturated wheel values are chosen using  Eq.~\eqref{eq:RW_rate_redistribution}.

    If $||\mbf{h}^{\text{RW}}_{4}||_\infty \leq h^{\text{RW}}_{\text{max}}$, then the allocation process is completed. Else, the process returns to Step 2.
    
\end{enumerate}

        Steps 2 through 4 of this process continue recursively until all components of the final wheel momentum $\mbf{h}^{\text{RW}}_4$ are within $\pm h_{\text{max}}$.
        Note that the initial pseudo-inverse solved in Step 1 determines an allocation with the smallest magnitude for the under-determined systems.
        When 1 RW saturates ($n = 3$), the reduced mapping becomes an one-to-one inverse mapping, leading to a unique solution in the allocation of the unsaturated wheels. 
        When 2 or 3 RWs are identified as saturated ($n \in \{1, 2\}$), the mapping becomes an over-determined system, and an exact solution does not exist. In this case, the allocations performed in Eqs.~\eqref{eq:RW_redistribution} and~\eqref{eq:RW_rate_redistribution} of Step 3 become a least-squares problems. A direct consequence of this is that the rate of change of the angular momentum in the 4 RWs, $\dot{\mbf{h}}^{\text{RW}}_{4}$, will not necessarily produce the desired value of $\dot{\mbf{h}}_{b,\text{des}}^{\text{RW}}$ from the attitude control in Eq.~\eqref{eq:PID_law}, and performance of the attitude controller may suffer. 
        When all RWs saturate, the 4-RW system can no longer provide any torque and attitude control is no longer possible. 
        

\subsection{Momentum Management Control Actuation} \label{sec:MM_actuation}

The momentum management time step is chosen based on the AMT position command update period of $\Delta t = 100$ seconds used on NEA Scout~\citep{orphee2018solar}, which is much longer than the attitude control (or ADCS) time step that is assumed to be one second in this work.
Specifically, the AMT and RCD control inputs generated by the momentum management controller are updated at every 100-second momentum management time step, while the numerical simulation of the sailcraft's nonlinear dynamics runs at a one-second resolution.

The AMT actuation input $\mbf{u}^\text{AMT}$ corresponds to the first two components of $\mbf{r}^{ps}_b$, where $\mbf{u}^\text{AMT} = \bbm r^{\text{AMT}}_{b1} & r^{\text{AMT}}_{b2} \ebm^\trans$.
At any given time $t_k$ for momentum management update, the AMT actuation dynamics are modeled based on the current position $\mbf{u}^\text{AMT}(t_k)$ and the incoming momentum management command $\mbf{u}^\text{AMT}_{k}$. For $t_k \leq t \leq t_k+\Delta t$, the system drives the AMT actuator along each axis ($i=1,2$) at its maximum speed $|\dot{u}^\text{AMT}_{\text{max},i}|$ toward the target command $u^\text{AMT}_{k,i}$. Once the commanded position is reached, the actuator stops and holds its position. These resulting transient position $\mbf{r}^{ps}_b(t)$ and velocity $\dot{\mbf{r}}^{ps}_b(t)$ are then used in the numerical simulation for the nonlinear system dynamics in Eq.~\eqref{EOM}.

\begin{figure}[b!]
    \centering
        \includegraphics[width=0.6\textwidth]{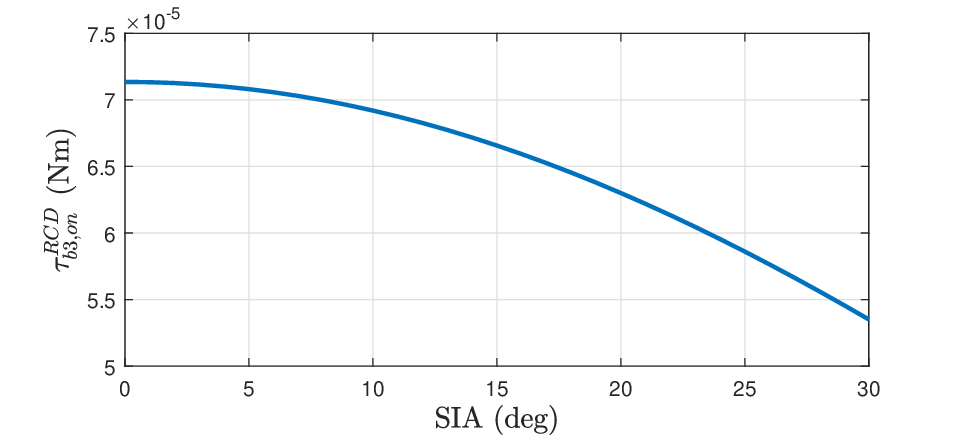}
    \caption{The RCD torque profile as functions of SIA.}\label{Fig:RCD_torque_profile}
\end{figure}

The layout of Solar Cruiser's RCDs allows for an approximately pure on-off roll momentum management torque to be generated in either direction~\citep{Inness2023MM,Tyler2024,heaton2023RCD}. Thus, the RCD torque is modeled as $\mbs{\tau}_b^{\text{RCD}} = \bbm 0 & 0 & \tau^\text{RCD}_{b3} \ebm ^\trans$, where $\tau^\text{RCD}_{b3} \in \{-\tau^\text{RCD}_{b3,\text{on}}, 0, \tau^\text{RCD}_{b3,\text{on}} \}$. 
The roll torque generated when the RCDs are turned on is set to meet the Solar Cruiser’s roll torque requirement of $6.525\times 10^{-5}$~N$\cdot$m at a $17^\circ$ sun incidence angle (SIA), which is $1.5$ times the sum of worst case roll disturbance and AMT induced roll torque at its maximum position offset~\citep{heaton2023RCD,johnson2022nasa}.
The RCD torque magnitude is modeled as a quadratic cosine function of SIA to match the analysis of~\cite{heaton2023RCD}, and is defined as
\beq \label{eq:RCD_torque}
\tau^\text{RCD}_{b3,\text{on}} = \f{6.525\times10^{-5}}{\cos^2(17^\circ)} \cos^2(\text{SIA}).
\eeq
The torque profile of $\tau^\text{RCD}_{b3,\text{on}}$ as a function of SIA is presented in Fig.~\ref{Fig:RCD_torque_profile} for a range of $0^\circ$ to $30^\circ$. This span covers roughly double the operational range of Solar Cruiser, which is designed to remain within an SIA of $0^\circ$ to $17^\circ$~\citep{Tyler2024,heaton2023RCD}.

\subsection{Linear Dynamics for Estimation and Prediction} \label{sec:LTVmodel}
For practical onboard implementation, a linear model is required for disturbance estimation and predictive control.
The accuracy of the model significantly affects the performance of the controller. However, higher model fidelity comes at the cost of increased computational demand. To enable real-time onboard implementation, a trade-off must be made between prediction accuracy and computational feasibility. 
Although a nonlinear dynamics model would have higher fidelity, it is not practical to consider the implementation of nonlinear MPC onboard with current technology due to their excessive computation demand.
A discrete-time linear model is thus used to supplement onboard disturbance estimation and predictive control, specifically the process model in Kalman filter and the prediction model in MPC.

The nonlinear dynamics model is linearized about the current state and AMT position at every time step, with knowledge of the attitude-dependent SRP force and estimated disturbance torque.
This yields a continuous-time linear time-varying (LTV) model, where the states include the attitude, angular velocity, reaction wheel angular momentum, and an integral state from the integral term of the attitude controller. 
The state of the linear system is denoted as $\mbf{x} = \bbm \mbs{\theta}^\trans & \mbs{\omega}^{ba^\trans}_b & \mbf{h}^{\text{RW}^\trans}_b & \mbf{e}^{\text{int}^\trans} \ebm^\trans$, where
$ \mbf{e}^\text{int} = \int^t_{t_0} \big(\mbs{\theta}(\tau) - \mbs{\theta}_d(\tau)\big) \dee\tau$ is the internal state representing the integral term of PID law.
It is assumed that the SRP force $\bar{\mbf{f}}_b^{\text{SRP}}$ is known through an estimate from the onboard ADCS. 
External disturbances are represented by $\mbf{w} = \mbs{\tau}^\text{dist}_b$, while $\mbf{u}^\text{AMT}$ and ${u}^\text{RCD}$ denote the AMT position and RCD torque input, respectively.
In order to perform time-varying trajectory tracking of $\mbs{\theta}_d$ and $\mbs{\omega}_d$, the system is linearized about the operation point of the desired trajectory with the current angular momentum and integral state $\bar{\mbf{x}} = \Big[ \mbs{\theta}_d^\trans \,\,\, \mbs{\omega}_d^{\trans} \,\,\, \bar{\mbf{h}}^{\text{RW}^\trans}_b \,\,\,  \mbfbar{e}^{\text{int}^\trans} \Big]^\trans$, the current AMT position $\mbfbar{r}^{ps}_b$, a nominally-off RCD torque $\bar{\mbs{\tau}}^\text{RCD}_{b} = \mbf{0}$, and the current SRP force $\bar{\mbf{f}}_b^{\text{SRP}}$, which are chosen as the current values when performing the linearization. Since future AMT positions are not known in advance, their rates are assumed to be zero in the linearization, \ie, $\dot{\bar{\mbf{r}}}_b^{ps} = \mbf{0}$, $\ddot{\bar{\mbf{r}}}_b^{ps} = \mbf{0}$.
The linearized continuous-time dynamics retaining the first-order term of Taylor series expansion is derived as
\beq
\dot{\mbf{x}} = \mbf{A}\mbf{x} + \mbf{B}_w\mbf{w} + \mbf{B}_{u1}\mbf{u}^\text{AMT} +  \mbf{B}_{u2}u^\text{RCD} +\mbf{z}, \label{eq:CT_LinEOM}
\eeq
where 
\bdis
\mbf{A} = \mbf{A}(\bar{\mbf{x}},\mbfbar{r}^{ps}_b) = \bbm \mbf{0}_{3 \times 3} &\mbf{1}_{3 \times 3} &\mbf{0}_{3 \times 3} & \mbf{0}_{3 \times 3} \\ 
    -\mbfbar{J}_b^{\mathcal{B}c^{-1}}\mbf{K}_{p} &\f{\p \mbf{f}_2}{\p\mbs{\omega}_b^{ba}}\Bigg|_{\bar{\mbf{x}}, \mbfbar{r}^{ps}_b} & -\mbfbar{J}_b^{\mathcal{B}c^{-1}}\bar{\mbs{\omega}}_b^{ba^\times} & -\mbfbar{J}_b^{\mathcal{B}c^{-1}}\mbf{K}_{i} \\ 
    \mbf{K}_{p} &\mbf{K}_{d} &\mbf{0}_{3 \times 3}&\mbf{K}_{i}  \\ 
        \mbf{1}_{3 \times 3}&\mbf{0}_{3 \times 3}&\mbf{0}_{3 \times 3}&\mbf{0}_{3 \times 3} \ebm,
\edis
\bdis
    \mbf{B}_{w} = \mbf{B}_{w}(\mbfbar{r}^{ps}_b) = \bbm \mbf{0}_{3 \times 3}\\ \mbfbar{J}_b^{\mathcal{B}c^{-1}} \\ \mbf{0}_{3 \times 3} \\\mbf{0}_{3 \times 3}  \ebm ,
    \hspace{1em} 
    \mbf{B}_{u1} = \mbf{B}_{u1}(\bar{\mbf{x}},\mbfbar{r}^{ps}_b,\mbfbar{f}_b^\text{SRP}) = \bbm \mbf{0}_{3 \times 2} \\ \f{\p \mbf{f}_2}{\p\mbf{r}_b^{ps}}\Bigg|_{\bar{\mbf{x}}, \mbfbar{r}^{ps}_b, \mbfbar{f}_b^\text{SRP}}\bbm \mbf{1}_{2 \times 2}\\ \mbf{0}_{1 \times 2} \ebm  \\ \mbf{0}_{3 \times 2} \\ \mbf{0}_{3 \times 2} \ebm,
    \hspace{1em} 
    \mbf{B}_{u2} = \mbf{B}_{u2}(\mbfbar{r}^{ps}_b) = \bbm \mbf{0}_{3 \times 3} \\ \mbfbar{J}_b^{\mathcal{B}c^{-1}} \\ \mbf{0}_{3 \times 3} \\ \mbf{0}_{3 \times 3} \ebm \bbm 0\\0\\1\ebm,
\edis
\bdis
\mbf{z} = \mbf{z}(\bar{\mbf{x}},\mbfbar{r}^{ps}_b,\bar{\mbs{\tau}}^\text{dist}_b,\mbfbar{f}_b^{\text{SRP}}, \bar{\mbs{\tau}}_b^{\text{dist}}) = \mbf{f}(\bar{\mbf{x}}, \mbfbar{r}^{ps}_b, \bar{\mbs{\tau}}^\text{RCD}_{b}, \mbfbar{f}_b^{\text{SRP}}, \bar{\mbs{\tau}}_b^{\text{dist}}) -\mbf{A}\bar{\mbf{x}} -\mbf{B}_w\bar{\mbs{\tau}}^\text{dist}_b -\mbf{B}_{u1}\mbfbar{r}^{ps}_b,
\edis
and
\begin{align*}
    &\f{\p \mbf{f}_2}{\p\mbs{\omega}_b^{ba}}\Big|_{\bar{\mbf{x}}, \mbfbar{r}^{ps}_b}  = \mbfbar{J}_b^{\mathcal{B}c^{-1}} \Big( \big(\mbfbar{J}_b^{\mathcal{B}c}\bar{\mbs{\omega}}_b^{ba}\big)^\times -\bar{\mbs{\omega}}_b^{ba^\times}\mbfbar{J}_b^{\mathcal{B}c} +\mbfbar{h}^{\text{RW}^\times}_b -\mbf{K}_d
        \Big), \\
    &\f{\p \mbf{f}_2}{\p\mbf{r}_b^{ps}}\Big|_{\bar{\mbf{x}}, \mbfbar{r}^{ps}_b,\mbfbar{f}_b^\text{SRP},\bar{\mbs{\tau}}_b^{\text{dist}}} = \mbfbar{J}_b^{\mathcal{B}c^{-1}} \Big( -\f{m_p^3+m_s^3}{(m_p+m_s)^2}\mbsbar{\omega}_b^{ba^\times} \Big(\mbfbar{r}^{ps^\times}_b\mbsbar{\omega}_b^{ba^\times} + \big( \mbfbar{r}^{ps^\times}_b \mbsbar{\omega}_b^{ba} \big)^\times \Big)  -\f{m_p^3+m_s^3}{(m_p+m_s)^2} \Big(\mbfbar{r}^{ps^\times}_b\dot{\mbsbar{\omega}}_b^{ba^\times} + \big( \mbfbar{r}^{ps^\times}_b\dot{\mbsbar{\omega}}_b^{ba}\big)^\times \Big) -\f{m_s}{m_p+m_s}\mbfbar{f}_b^{\text{SRP}^\times} \Big).
\end{align*}
Note that the state-space matrices depend on $\mbfbar{r}^{ps}_b$ because $\mbf{J}_b^{\mathcal{B}c}$ is a function of $\mbf{r}^{ps}_b$. The nonlinear function is defined as
\bdis
\mbf{f}(\bar{\mbf{x}}, \mbfbar{r}^{ps}_b, \bar{\mbs{\tau}}^\text{RCD}_{b}, \mbfbar{f}_b^{\text{SRP}}, \bar{\mbs{\tau}}_b^{\text{dist}}) = \bbm \mbf{S}^{-1}(\mbs{\theta}_d) \\ \mbf{f}_2(\bar{\mbf{x}}, \mbfbar{r}^{ps}_b, \bar{\mbs{\tau}}^\text{RCD}_{b}, \mbfbar{f}_b^{\text{SRP}}, \bar{\mbs{\tau}}_b^{\text{dist}}) \\ \mbf{K}_i \bar{\mbf{e}}^\text{int} \\ \mbf{0} \ebm,
\edis
where
\begin{align*}
    \mbf{f}_2 = \dot{\mbs{\omega}}_b^{ba} = \mbf{J}_b^{{\mathcal{B}c}^{-1}} \Bigg( &-\mbs{\omega}_b^{ba^\times} \mbf{J}_b^{\mathcal{B}c} \mbs{\omega}_b^{ba} 
    -\mbs{\omega}_b^{ba^\times}\mbf{h}_b^{\text{RW}} 
    -\f{m_p^3+m_s^3} {(m_p+m_s)^2} \bigg( \mbf{r}_b^{ps^\times}\ddot{\mbf{r}}_b^{ps} -2\dot{\mbf{r}}_b^{ps^\times} \mbf{r}_b^{ps^\times} \mbs{\omega}_b^{ba} \bigg) \\
    &+\f{m_s}{m_p+m_s}\mbf{r}_b^{ps^\times} \mbf{f}_b^{\text{SRP}} 
    + \mbs{\tau}_b^{\text{RCD}} 
    + \mbs{\tau}_b^{\text{dist}} 
    -\mbf{K}_p (\mbs{\theta}-\mbs{\theta}_d) -\mbf{K}_d (\mbs{\omega}^{ba}_b-\mbs{\omega}_d) -\mbf{K}_i \int^t_{t_0} \bigg(\mbs{\theta}(\tau) - \mbs{\theta}_d(\tau)\bigg) \dee\tau  \Bigg) 
\end{align*} 
is a combination of the attitude dynamics in Eq.~\eqref{EOM} and the RW control law in Eq.~\eqref{eq:PID_law} that characterize $\dot{\mbs{\omega}}_b^{ba}$.

In this work, a zeroth-order hold (ZOH) discretization on both AMT and RCD actuation is used, which has a lower computation requirement when compared to the mixed-FOH-ZOH discretization employed by~\cite{shenISSS2025}. 
Discretizing Eq.~\eqref{eq:CT_LinEOM} using a ZOH with the momentum management timestep $\Delta t$ results in 
\beq
    \mbf{x}_{k} = \mbf{A}_{k}\mbf{x}_{k} + \mbf{B}_{w,k}\mbf{w}_{k} + \mbf{B}_{u1,k}\mbf{u}^{\text{AMT}}_{k} + \mbf{B}_{u2,k}{u}^{\text{RCD}}_{k} +\mbf{z}_k, \label{eq:DT_dynamics_ss}
\eeq
which is used as the the Kalman filter process model and MPC prediction model. 
The discrete-time LTV state-space matrices are obtained by solving $\mbf{A}_{k} = \mbs{\Phi}(t_{k+1},t_k)$, $\bbm \mbf{B}_{w,k} & \mbf{B}_{u1,k} & \mbf{B}_{u2,k}\ebm = \mbf{A}_{k}\tilde{\mbf{B}}_{k}$, and $\mbf{z}_k = \mbf{A}_{k}\tilde{\mbf{z}}_{k}$ through numerical integration of the matrix differential equations
\begin{align*}
    &\dot{\mbs{\Phi}}(t,t_k) = \mbf{A}\mbs{\Phi}(t,t_k),\\
    &\dot{\tilde{\mbf{B}}}_{k}(t,t_k) = \mbs{\Phi}^{-1}(t,t_k)\bbm \mbf{B}_{w} & \mbf{B}_{u1} & \mbf{B}_{u2}\ebm,\\
    &\dot{\tilde{\mbf{z}}}_{k}(t,t_k) = \mbs{\Phi}^{-1}(t,t_k) \mbf{z},
\end{align*}
with the initial values $\mbs{\Phi}(t_{k+1},t_k) = \mbf{1}$, $\tilde{\mbf{B}}_{k}(t_k) = \mbf{0}$, $\tilde{\mbf{z}}_{k}(t_k) = \mbf{0}$ over the time interval $t \in [t_k, t_{k+1}]$.
Note that the nonlinear dynamics in Eq.~\eqref{EOM} are used as the system's dynamics for all numerical simulations, while the Kalman filter and MPC use the discrete-time LTV model.
Section~\ref{sec:KF} presents a state and disturbance estimation framework based on a Kalman filter that is used to yield the estimates $\hat{\mbf{x}}(t_k)$ and $\hat{\mbf{w}}(t_k)$ needed to compute the LTV dynamics model used in MPC.


\section{Disturbance Estimation Using Kalman Filter} \label{sec:KF}

Given NASA's efforts to develop on-orbit SRP calibration algorithms~\citep{carzana2023solar}, it is reasonable to assume that an accurate SRP force estimate is available within the proposed momentum management algorithm. However, the SRP disturbance torque cannot be measured directly, and needs to be estimated onboard in real time. The disturbance torque $\mbs{\tau}^\text{dist}_b$ is an external input acting on the solar sail system that impacts its dynamics.
In the MPC policy proposed in Section~\ref{sec:MPC}, this disturbance torque is a key parameter in the prediction model used to forecast the system dynamics and determine optimal momentum management actuation.
Due to the slow motion and relatively steady attitude operation nature of solar sails, the disturbance torque is modeled as approximately constant or slow varying.
A Kalman filter framework is developed in this section to supplement this essential parameter for MPC momentum management.

\subsection{Measurement Model}

Solar Cruiser's ADCS provides an accurate estimate of the sailcraft's attitude, angular velocity, and angular momentum using onboard sensors such as rate gyros, inertial measurement units (IMUs), sun sensors, and star trackers.
It is thus assumed in this work that a full state measurement of $\mbf{x}_k = \bbm \mbs{\theta}^\trans_k & \mbs{\omega}^{ba^\trans}_{b,k} & \mbf{h}^{\text{RW}^\trans}_{b,k} & \mbf{e}^{\text{int}^\trans}_k \ebm^\trans$ is accessible, and the measurement noise is normally distributed, resulting in the measurement model
\bdis
    \mbf{y}_{k} = \underbrace{\bbm \mbf{1}_{12\times12} & \mbf{0}_{12\times3} \ebm}_{\mbf{H}} \bbm \mbf{x}_k \\ \mbf{w}_k \ebm + \mbs{\nu}_{k}, \quad \mbs{\nu}_{k} \sim \mathcal{N}(\mbf{0}, \mbf{R}^\text{KF}) ,
\edis
where $\mbs{\nu}_{k}$ is the linear additive measurement noise, and the measurement error covariance matrix $\mbf{R}^\text{KF} = \text{diag}(\mbf{r}_\theta, \mbf{r}_\omega, \mbf{r}_h, \mbf{r}_e)$ is determined by the variance of each corresponding state measurement error ($\mbs{\sigma}_\theta^2, \mbs{\sigma}_\omega^2, \mbs{\sigma}_h^2, \mbs{\sigma}_e^2$), which is associated with the onboard ADCS state estimation accuracy. 

\subsection{Process Model}
Given the slow evolution of the spacecraft attitude, it is assumed that SRP force $\mbfbar{f}_b^{\text{SRP}}$ is a known constant updated at every time step, the error of the dynamics model is Gaussian and linearly additive, and the disturbance torque to be estimated, $\mbs{\tau}_b^{\text{dist}}$, is constant within the time update (prediction) step.
The discrete-time LTV model in Eq.~\eqref{eq:DT_dynamics_ss} with the addition of linear model error is given by
\begin{align*}
    \mbf{x}^\text{KF}_{k+1} &= \mbf{A}_k \mbf{x}^\text{KF}_k + \mbf{B}_{w,k} \mbf{w}^\text{KF}_k + \mbf{B}_{u1,k} \mbf{u}^\text{AMT}_k + \mbf{B}_{u2,k} {u}^\text{RCD}_{k} + \mbf{z}_k + \mbs{\eta}^\text{model}_k, \quad \mbs{\eta}^\text{model}_k \sim \mathcal{N}(\mbf{0}, \mbf{Q}^\text{KF}_\text{model}), \\   
    \mbf{w}^\text{KF}_{k+1} &= \mbf{w}^\text{KF}_{k} + \mbs{\eta}^\text{dist}_k, \quad \mbs{\eta}^\text{dist}_k \sim \mathcal{N}(\mbf{0}, \mbf{Q}^\text{KF}_\text{dist}), 
\end{align*}
where the linear model error $\mbs{\eta}^\text{model}_k$ and disturbance error $\mbs{\eta}^\text{dist}_k$ are assumed to be linearly additive and normally distributed.
The error covariance matrices of the model uncertainty are $\mbf{Q}^\text{KF}_\text{model} = \text{diag}(\mbf{q}_\theta, \mbf{q}_\omega, \mbf{q}_h, \mbf{q}_e)$ and $\mbf{Q}^\text{KF}_\text{dist}$, which are tuning parameters chosen to influence the Kalman filter's aggressiveness in updating the estimate of the disturbance torque $\mbf{w}^\text{KF}$. In particular, increasing the variances within $\mbf{Q}^\text{KF}_\text{model}$ results in a slower convergence of the disturbance torque, while increase the variances within $\mbf{Q}^\text{KF}_\text{dist}$ speeds up the convergence, while potentially making the estimates more sensitive to measurement noise.
In order to adapt to the attitude-dependent disturbance torque varying with the slew maneuver, a dynamic error covariance scaling quadratically with the desired angular velocity is added to the disturbance covariance matrix, such that $\mbf{Q}^\text{KF}_\text{dist} = \text{diag}( {q}_{\tau, 1}+\xi_1\omega_{d,1}^2, {q}_{\tau,2}+\xi_2\omega_{d,2}^2, {q}_{\tau,3} + \xi_3 \omega_{d,3}^2)$, and $\xi_i$ is the scaling parameter associate to the dynamic covariance for $i=1,2,3$.

The complete Kalman filter process model is reformulated as
\bdis
\underbrace{\bbm \hat{\mbf{x}}^-_{k+1} \\ \hat{\mbf{w}}^-_{k+1} \ebm}_{\hat{\mbf{X}}^-_{k+1}} 
= \underbrace{\bbm \mbf{A}_k & \mbf{B}_{w,k} \\ \mbf{0}_{3\times12} & \mbf{1}_{3\times3} \ebm}_{\mbf{F}_k} \underbrace{\bbm \hat{\mbf{x}}^+_{k} \\ \hat{\mbf{w}}^+_k \ebm}_{\hat{\mbf{X}}^+_k} + \underbrace{\bbm \mbf{B}_{u,k} \\ \mbf{0}_{3\times3} \ebm}_{\mbf{G}_k} \underbrace{\bbm \mbf{u}^\text{AMT}_k \\ {u}^\text{RCD}_{k} \ebm}_{\mbf{U}_k} +\underbrace{\bbm \mbf{z}_k \\ \mbf{0} \ebm}_{\mbf{Z}_k} + \underbrace{\bbm \mbs{\eta}^\text{model}_k \\ \mbs{\eta}^\text{dist}_k \ebm}_{\mbs{\eta}_k} ,
\edis
where the process noise is given by $\mbs{\eta}_k \sim \mathcal{N}(\mbf{0}, \mbf{Q}^\text{KF} )$ and $\mbf{Q}^\text{KF} = \text{diag}(\mbf{Q}^\text{KF}_\text{model},\mbf{Q}^\text{KF}_\text{dist})$.

\subsection{Summary of Kalman Filter Estimation Framework}

In the time update (prediction) step, the a priori (predicted) state estimate and error covariance are given by
\begin{align*}
    \hat{\mbf{X}}_k^- &= \mbf{F}_{k-1} \hat{\mbf{X}}_{k-1}^+ + \mbf{G}_{k-1} \mbf{U}_{k-1} +\mbf{Z}_{k-1}, \\
    \mbf{P}_k^- &= \mbf{F}_{k-1} \mbf{P}_{k-1}^+ \mbf{F}_{k-1}^\trans + \mbf{Q}^\text{KF} .
\end{align*}
In the measurement update (correction) step, the a posteriori (updated) state estimate and error covariance are given by
\begin{align*}
    \hat{\mbf{X}}_k^+ &= \hat{\mbf{X}}_k^- + \mbf{K}_k \left( \mbf{y}_k - \mbf{H} \hat{\mbf{X}}_k^- \right), \\
    \mbf{P}_k^+ &= \left( \mbf{1} - \mbf{K}_k \mbf{H} \right) \mbf{P}_k^- ,
\end{align*}
where the Kalman gain is computed as $\mbf{K}_k = \mbf{P}_k^- \mbf{H}^\trans \left( \mbf{H} \mbf{P}_k^- \mbf{H}^\trans + \mbf{R}^\text{KF} \right)^{-1}.$ The state estimate $\hat{\mbf{X}}_k^+ = \bbm \hat{\mbf{x}}^{+^\trans}_{k} & \hat{\mbf{w}}^{+^\trans}_{k} \ebm^\trans $ is used within the momentum management controller presented in the following section.


\section{Momentum Management Using MPC} \label{sec:MPC}

Solar sail slew maneuvers are inherently slow due to the small magnitude of SRP torques and the sailcraft's large moment of inertia. As a result, the system dynamics are relatively smooth and predictable, and external disturbances such as SRP imbalance or environmental torques evolve gradually. Moreover, the long time scales involved in solar sail maneuvers provide sufficient computational time for onboard optimization. These characteristics make MPC particularly suitable for solar sail momentum management, where coordinated use of RWs and momentum management actuators (AMT and RCDs) is required to prevent RW saturation while maintaining accurate attitude control. This section presents the MPC framework tailored for solar sail momentum management, specifically designed for Solar Cruiser's configuration.


\subsection{Introduction to MPC}

MPC is an advanced optimal control strategy that computes control actions by solving a constrained optimization problem over a finite prediction horizon at each time step. It determines a sequence of control inputs that minimize a specified objective function while satisfying the system dynamics, actuator limits, and state constraints. 
At each control update, MPC uses the current system state and a predictive model to forecast future behavior over a finite horizon of $N$ time steps. It then solves for the optimal sequence of control inputs, yet only the first input is applied to the system. At the next time step, the process is repeated using updated measurements and system information. This receding-horizon strategy enables continual adaptation to disturbances and modeling inaccuracies, providing robust feedback control in the presence of uncertainty.

Real-time implementation of MPC onboard a flight computer can be realistically achieved by formulating the optimization problem as a convex QP with a quadratic objective function and affine constraints. The use of a linear dynamic prediction model within the MPC framework is required in order for it to be formulated as a QP.

\subsection{RCD Constraint Relaxation}
The control input $u^\text{RCD} = \tau^\text{RCD}_{b3}$ is chosen based on the momentum management strategy. 
Considering the RCD on-off actuation as an explicit constraint in the MPC optimization problem leads to a mixed-integer problem, which is computationally expensive, and limits the practicality of onboard real-time MPC.
To enable the use of convex optimization solvers with the proposed MPC policy, the integer constraint is relaxed, and a PWM quantization is applied to the RCD actuation~\citep{shen2025}.
A continuous value of $u^\text{RCD}_\text{mpc}$ is allowed in the optimization problem, where $-\tau^\text{RCD}_{b3,\text{on}} \leq u^\text{RCD}_\text{mpc} \leq \tau^\text{RCD}_{b3,\text{on}}$ and $\tau^\text{RCD}_{b3,\text{on}}$ is the roll torque magnitude generated when the RCDs are turned on.
After solving the MPC optimization problem, the continuous $u^\text{RCD}_\text{mpc}$ is then quantized into a discrete value 
\beq \label{eq:PWM_cases}
u^\text{RCD}(t) = 
\begin{cases}
\beta_\text{on}\tau^\text{RCD}_{b3,\text{on}}, \quad \text{for} \quad t_k \leq t < (t_k + t_c),\\
0, \quad \text{for} \quad (t_k + t_c) \leq t < t_{k+1},
\end{cases}
\eeq
where $\beta_\text{on} \in \{-1, 1\}$ denotes the clockwise and counterclockwise directions about the roll ($\vect{b}^3$) axis, $\tau^\text{RCD}_{b3,\text{on}}$ denotes the torque magnitude when RCDs are turned on, and $t_c= \Delta t \cdot \f{u^\text{RCD}_{\text{mpc}}}{\tau^\text{RCD}_{b3,\text{on}}}$ is the length (cut-off time) of a single pulse PWM conversion from a continuous MPC optimal RCD input.
Details of the RCD quantization can be found in the work of~\cite{shen2025}.


\subsection{Prediction Model}

In contrast to the state $\mbf{x}$ used in Section~\ref{sec:LTVmodel}, a modified state $\mbf{x}^\text{MPC} = \bbm \mbs{\theta}^\trans & \mbs{\omega}^{ba^\trans}_b & \mbf{h}^{\text{RW}^\trans}_4 & \mbf{e}^{\text{int}^\trans} \ebm^\trans$ is employed in this MPC framework, where $\mbf{h}^\text{RW}_4 = \mathbf{M}_{34}^\dagger \mbf{h}_b^{\text{RW}}$ follows the pseudo-inverse relationship discussed in Section~\ref{sec:RW}.
This modification allows for a direct constraint on the angular momentum of the individual RWs within the MPC framework.
Although a simple pseudo-inverse mapping is used in this prediction model, the proposed MPC approach is not limited to this specific optimal allocation method. More advanced RW angular momentum allocation synthesis can be used based on the design of the ADCS.

The linearized dynamics in Eq.~\eqref{eq:CT_LinEOM} are modified to obtain a linear prediction model to be used in the MPC framework. Specifically, the linear transformation $\mbf{x} = \mbf{T}\mbf{x}^\text{MPC}$ is applied, where $\mbf{T} = \text{diag}(\mbf{1},\mbf{1},\mbf{M}_{34},\mbf{1})$ is formed using the RW geometry matrix $\mbf{M}_{34}$.  The inverse linear transformation $\mbf{x}^\text{MPC} = \mbf{T}^\dagger\mbf{x}$ is computed using the pseudo-inverse of $\mbf{M}_{34}$ as $\mbf{T}^\dagger = \text{diag}(\mbf{1},\mbf{1},\mbf{M}_{34}^\dagger,\mbf{1})$.  Applying these transformations to Eq.~\eqref{eq:CT_LinEOM} yields the linear dynamics
\beq
\dot{\mbf{x}}^\text{MPC} = \mbf{A}^\text{MPC}\mbf{x}^\text{MPC} + \mbf{B}_w^\text{MPC}\mbf{w} + \mbf{B}_{u1}^\text{MPC}\mbf{u}^\text{AMT} +  \mbf{B}_{u2}^\text{MPC}u^\text{RCD} + \mbf{z}^\text{MPC}, \label{eq:CT_LinEOM_MMPC}
\eeq
where $\mbf{A}^\text{MPC} = \mbf{T}^\dagger\mbf{A}\mbf{T}$, $\mbf{B}_w^\text{MPC} = \mbf{T}^\dagger\mbf{B}_w$, $\mbf{B}_{u1}^\text{MPC} = \mbf{T}^\dagger\mbf{B}_{u1}$, $\mbf{B}_{u2}^\text{MPC} = \mbf{T}^\dagger\mbf{B}_{u2}$, and $\mbf{z}^\text{MPC} = \mbf{T}^\dagger\mbf{z}$. A ZOH is then applied to the inputs of Eq.~\eqref{eq:CT_LinEOM_MMPC} to yield the discrete-time linear prediction used by MPC over its prediction model, given by
\beq \label{eq:MPC_pred_model}
    \mbf{x}^{\text{MPC}}_{j+1|t_k} = \mbf{A}^{\text{MPC}}_{j|t_k}\mbf{x}^{\text{MPC}}_{j|t_k} + \mbf{B}^{\text{MPC}}_{w,j|t_k}\mbf{w}_{j|t_k} + \mbf{B}^{\text{MPC}}_{u1,j|t_k}\mbf{u}^{\text{AMT}}_{j|t_k} + \mbf{B}^{\text{MPC}}_{u2,j|t_k}{u}^{\text{RCD}}_{j|t_k}  + \mbf{z}^\text{MPC}_{j|t_k}, \quad j=0, 1, \ldots, N-1,  
\eeq
where the subscript $j|t_k$ refers to the $j$-th discrete time step within the MPC prediction horizon at time step $t_k$. 
The prediction model includes the state $\mbf{x}^{\text{MPC}}_{j|t_k} =  \bbm \mbs{\theta}_{j|t_k}^\trans & \mbs{\omega}^{ba^\trans}_{b,j|t_k} & \mbf{h}_{4,j|t_k}^{\text{RW}^\trans} & \mbf{e}_{j|t_k}^{\text{int}^\trans} \ebm^\trans$, the AMT input $\mbf{u}^{\text{AMT}}_{j|t_k}$, and the RCD input ${u}^{\text{RCD}}_{j|t_k}$ to be designed across the prediction horizon $j=0, 1, \ldots, N-1$ at time $t_k$. The disturbance torque is represented by $\mbf{w}_{j|t_k}$. The MPC framework uses the state estimate from Kalman filter framework presented in Section~\ref{sec:KF} as its knowledge within the prediction model, where $\mbf{x}^{\text{MPC}}_{0|t_k} =  \bbm \hat{\mbs{\theta}}_k^\trans & \hat{\mbs{\omega}}^{ba^\trans}_{b,k} & \mathbf{M}_{34}^\dagger\hat{\mbf{h}}_{b,k}^{\text{RW}^\trans} & \hat{\mbf{e}}_k^{\text{int}^\trans} \ebm^\trans$ and $\mbf{w}_{j|t_k} = \hat{\mbf{w}}^+_{k}$ for $j=0, 1, \ldots, N-1$.
The $\hat{(\cdot)}^+$ notation for a posteriori (measurement updated) state estimate is simplified as $\hat{(\cdot)}$ to avoid notation clustering.
The LTV matrices of $\mbf{A}^{\text{MPC}}_{j|t_k}$, $\mbf{B}^{\text{MPC}}_{w,j|t_k}$, $\mbf{B}^{\text{MPC}}_{u1,j|t_k}$, $\mbf{B}^\text{MPC}_{u2,j|t_k}$, $\mbf{z}^\text{MPC}_{j|t_k}$ are linearized about the operation points (varying with the slew maneuver) across the prediction horizon, where $\bar{\mbf{x}}^{\text{MPC}}_{j|t_k} = \Big[ \mbs{\theta}_{d,j|t_k}^\trans \,\,\, \mbs{\omega}_{d,j|t_k}^{\trans} \,\,\, \mathbf{M}_{34}^\dagger\hat{\mbf{h}}_{b,k}^{\text{RW}^\trans} \,\,\,  \hat{\mbf{e}}^{\text{int}^\trans}_k \Big]^\trans$ for $j=0, 1, \ldots, N-1$, $\mbfbar{r}^{ps}_{b} = \mbfbar{r}^{ps}_b(t_k)$, $\bar{\mbf{f}}_{b}^{\text{SRP}} = \mbf{f}_b^{\text{SRP}}(t_k)$ are updated at every momentum management timestep at time $t_k$, and $\dot{\bar{\mbf{r}}}_{b}^{ps} = \ddot{\bar{\mbf{r}}}_{b}^{ps} = \mbf{0}$, $\bar{\mbs{\tau}}^\text{RCD}_{b} = \mbf{0}$. The time-varying parameters are the desired attitude $\mbs{\theta}_{d,j|t_k}$ and desired angular velocity $\mbs{\omega}_{d,j|t_k}$, while the SRP force, disturbance torque, state (angular momentum and integral state), and AMT position to be linearized about are kept constant throughout the MPC prediction horizon. 


\subsection{State Constraints}

To ensure the practical feasibility of the controller, inequality constraints are imposed on the system states. These constraints enforce bounded deviations in sailcraft attitude, angular velocity, RW angular momentum, and integrated attitude error across the prediction horizon as $\mbs{\theta}_{d,j|t_k} -\mbs{\theta}_\text{err} \leq \mbs{\theta}_{j|t_k} \leq \mbs{\theta}_{d,j|t_k} +\mbs{\theta}_\text{err}$, $\mbs{\omega}_{d,j|t_k} -\mbs{\omega}_\text{err} \leq \mbs{\omega}_{b,j|t_k}^{ba} \leq \mbs{\omega}_{d,j|t_k} +\mbs{\omega}_{\text{err}}$, $\mbf{h}^\text{RW}_{4,\text{min}} \leq \mbf{h}_{4,j|t_k}^\text{RW} \leq \mbf{h}^\text{RW}_{4,\text{max}}$, and $\mbf{e}^\text{int}_\text{min} \leq \mbf{e}^\text{int}_{j|t_k} \leq \mbf{e}^\text{int}_\text{max}$, respectively. The variables $\mbs{\theta}_\text{err}$ and $\mbs{\omega}_\text{err}$ are the allowable deviation from the nominal slew trajectory $\mbs{\theta}_{d,j|t_k}$ and $\mbs{\omega}_{d,j|t_k}$ across the prediction horizon. Collectively, this is written as the constraint $\mbf{x}^{\text{MPC}}_{\text{min}} \leq \mbf{x}^{\text{MPC}}_{j|t_k} \leq \mbf{x}^{\text{MPC}}_{\text{max}}$.

To further avoid the RW angular momentum approaching the hardware physical saturation limits during sustained disturbance rejection accommodate modeling errors, soft constraints are introduced to incentivize the RW angular momentum to stay within a safe operational margin. 
These soft bounds are defined as 
$$
    \mbf{h}_{4,\text{min}}^\text{soft} - \mbs{\alpha} \leq \mbf{h}^\text{RW}_{4,j|t_k} \leq  \mbf{h}_{4,\text{max}}^\text{soft} + \mbs{\alpha},
$$
where where $\mbf{h}_{4,\text{min}}^\text{soft}$ and $\mbf{h}_{4,\text{max}}^\text{soft}$ represent the lower and upper bounds of the soft constraint envelope (i.e., the desired safe operation region), and the non-negative slack variable $\mbs{\alpha} \geq \mbf{0}$ is quadratically penalized in the MPC objective function, enabling graceful constraint relaxation while encouraging the system to remain within the nominal safe range.
Within the soft bounds, the slack variable remains zero and no penalty is incurred. When violated, the controller attempts to reduce the non-zero value of $\mbs{\alpha}$ to drive the RWs angular momentum back within the safe region, avoiding saturation.

\begin{figure}[b!]
    \centering
        \includegraphics[width=0.6\textwidth]{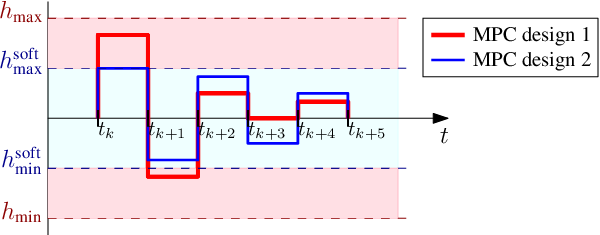}
    \caption{Illustration of the MPC soft constraint design with a prediction horizon of $N=5$, where no penalty is incurred for responses satisfying $h^\text{soft}_\text{min} \leq h \leq h^\text{soft}_\text{max}$, while a quadratic penalty appears in the MPC objective function when $h^\text{soft}_\text{max} \leq h$ or $h \leq h^\text{soft}_\text{min}$. The response labeled ``MPC design 1'' indicates a design that violates the soft constraint, while ``MPC design 2'' does not.}\label{Fig:MPC_soft}
\end{figure}

Figure~\ref{Fig:MPC_soft} illustrates the relationship between the soft constraint and the operational limits of the RWs. The original hard limits $h_\text{max}$ and $h_\text{min}$ define the absolute, physically-imposed boundaries that cannot be violated. The soft constraint bounds $h^\text{soft}_\text{max}$ and $h^\text{soft}_\text{min}$ define the preferred operating limits. The light blue area defined by $h^\text{soft}_\text{min} \leq h \leq h^\text{soft}_\text{max}$ is the region where the soft constraint is satisfied and the slack variable $\mbs{\alpha}$ is zero and has not effect on the MPC objective function. The light red area defined by $h^\text{soft}_\text{max} \leq h$ or $h \leq h^\text{soft}_\text{min}$ is the region where the soft constraint is violated. When the MPC design variable enters this region, the slack variable $\mbs{\alpha}$ takes on a positive value and the violation is heavily penalized in the objective function. Two examples of design variable sequence interpreting the design choices in the MPC optimization are shown in Fig.~\ref{Fig:MPC_soft}. The red trajectory (labeled as ``MPC design 1'') represents an action that violates the soft constraint in the first two steps, incurring a large penalty due to the quadratic weight on the slack variable in the MPC objective function. The blue trajectory (labeled as ``MPC design 2'') represents a sequence of design that remains within the feasible region, incurring no penalty within the MPC objective function. 

The soft constraint serves to improve feasibility of the MPC optimization problem by penalizing, rather than prohibiting, constraint violation. This structure strongly discourages the design variables from exceeding the soft bounds $h^\text{soft}_\text{max} \leq h \leq h^\text{soft}_\text{min}$, but allows for excursions outside this region if the performance benefit outweighs the imposed penalty.


\subsection{Input Constraints}

To ensure actuator feasibility and the satisfaction of hardware limits, constraints are imposed on both the control input magnitude and the rate of AMT motion. The actuator input $\mbf{u}_{j|t_k} = \bbm \mbf{u}^{\text{AMT}^\trans}_{j|t_k} &{u}^{\text{RCD}}_{j|t_k}\ebm ^\trans$ is subject to the constraints
$\mbf{u}_{\text{min}} \leq \mbf{u}_{j|t_k} \leq \mbf{u}_{\text{max}}$, 
where the bounds $\mbf{u}_{\text{min}} = \bbm \mbf{u}^{\text{AMT}^\trans}_{\text{min}} & -\tau^\text{RCD}_{b3,\text{on}}\ebm ^\trans$ and $\mbf{u}_{\text{max}} = \bbm \mbf{u}^{\text{AMT}^\trans}_{\text{max}} & \tau^\text{RCD}_{b3,\text{on}}\ebm ^\trans$ reflect the physical actuation limits of the AMT position and RCD torque. It is assumed that the RCD torque magnitude is accessible from the ADCS, which can be computed based on the SIA.

In addition, the translational rate of the AMT is constrained to avoid unrealistic or dynamically infeasible actuation commands, referring to the physical speed limit of AMT. Due to the discrete-time formulation of the MPC problem, the rate constraint is implemented as a finite difference inequality 
$$
\dot{\mbf{u}}^\text{AMT}_{\text{min}} \leq \frac{\mbf{u}^\text{AMT}_{j|t_k} - \mbf{u}^\text{AMT}_{j-1|t_k}}{\Delta t} \leq \dot{\mbf{u}}^\text{AMT}_{\text{max}},
$$
where $\dot{\mbf{u}}^\text{AMT}_{\text{min}}$ and $\dot{\mbf{u}}^\text{AMT}_{\text{max}}$ define the allowable lower and upper bounds on the AMT velocity.

To ensure input continuity across successive control intervals, which is critical for the AMT input between discrete time steps, the initial AMT input at each new MPC update must match the current AMT position. This continuity constraint is enforced as
$\mbf{u}^{\text{AMT}}_{-1|t_k} = \mbf{u}^{\text{AMT}}(t_k)$, where the current AMT position becomes a design variable fixed by this equality constraint and is used to constrain the AMT rate of the first input within the MPC optimization problem.
This formulation ensures smooth AMT motion while preserving the predictive accuracy and numerical stability of the MPC framework.


\subsection{Objective Function}

The objective function of the proposed MPC policy is formulated to balance state regulation, actuator efficiency, AMT motion minimization, and enforcement of soft constraints. It is defined as
\bdis
    \sum_{j=0}^{N-1} \Bigl{(} (\mbf{x}_{j|t_k}^\text{MPC}-\mbfbar{x}_{j|t_k}^\text{MPC})^\trans \mbf{Q} (\mbf{x}^\text{MPC}_{j|t_k}-\mbfbar{x}_{j|t_k}^\text{MPC}) + \mbf{u}_{j|t_k}^\trans \mbf{R} \mbf{u}_{j|t_k} + \tilde{\mbf{u}}^{\text{AMT}^\trans}_{j|t_k} \tilde{\mbf{R}}\tilde{\mbf{u}}^\text{AMT}_{j|t_k}\Bigl{)} 
    + (\mbf{x}_{N|t_k}^\text{MPC}-\mbfbar{x}_{N|t_k}^\text{MPC})^\trans \mbf{Q}_N (\mbf{x}^\text{MPC}_{N|t_k}-\mbfbar{x}_{N|t_k}^\text{MPC}) + \mbs{\alpha}^\trans\mbf{C}\mbs{\alpha} ,
\edis
where 
$\mbf{x}^\text{MPC}_{j|t_k}$ and $\mbf{u}_{j|t_k}$ denote the predicted state and control input at stage $j$ over the prediction horizon of length $N$; $\bar{\mbf{x}}^{\text{MPC}}_{j|t_k} = \Big[ \mbs{\theta}_{d,j|t_k}^\trans \,\,\, \mbs{\omega}_{d,j|t_k}^{\trans} \,\,\, \mathbf{M}_{34}^\dagger\hat{\mbf{h}}_{b,k}^{\text{RW}^\trans} \,\,\,  \hat{\mbf{e}}^{\text{int}^\trans}_k \Big]^\trans$ is the operating point at stage $j$ over the prediction horizon of length $N$;
$\mbf{Q} = \mbf{Q}^\trans$ and $\mbf{R} = \mbf{R}^\trans$ are positive semi-definite and positive definite weighting matrices, respectively, penalizing the state and control input;
$\mbf{Q}_N$ is the terminal weighting matrix for the final predicted state at stage $N$;
$\mbf{C} = \mbf{C}^\trans$ is a positive semi-definite matrix that penalizes violation of the soft constraint via the slack variable $\mbs{\alpha} \geq \mbf{0}$;
$\mbftilde{R} = \mbftilde{R}^\trans$ is a positive semi-definite matrix penalizing the rate of AMT translation; and $\tilde{\mbf{u}}^\text{AMT}_{j|t_k} = \mbf{u}^\text{AMT}_{j|t_k}-\mbf{u}^\text{AMT}_{j-1|t_k}$ is the difference between the $j$-th AMT input and the previous input.

The term $\tilde{\mbf{u}}^{\text{AMT}^\trans}_{j|t_k} \tilde{\mbf{R}}\tilde{\mbf{u}}^\text{AMT}_{j|t_k}$ is included to discourage unnecessary movement of the AMT, thereby promoting actuator efficiency and helping maintain the AMT in a relatively stationary configuration across time steps. This is particularly important given the discretized AMT inputs and the associated mechanical and dynamic constraints.
The slack variable penalty $\mbs{\alpha}^\trans \mbf{C} \mbs{\alpha}$ enables soft constraint enforcement on RW angular momentum, where violations are permitted when necessary, but discouraged through a quadratic penalty.


\subsection{Summary of MPC Policy and Implementation Details}

The proposed MPC policy involves solving for the optimization problem
\begin{align}
    &\minimize_{\mbs{\mathcal{X}},\hspace{2pt} \mbs{\mathcal{U}},\hspace{2pt} \mbs{\alpha}} \sum_{j=0}^{N-1} \Bigl{(} (\mbf{x}_{j|t_k}^\text{MPC}-\mbfbar{x}_{j|t_k}^\text{MPC})^\trans \mbf{Q} (\mbf{x}^\text{MPC}_{j|t_k}-\mbfbar{x}_{j|t_k}^\text{MPC}) + \mbf{u}_{j|t_k}^\trans \mbf{R} \mbf{u}_{j|t_k} + \tilde{\mbf{u}}^{\text{AMT}^\trans}_{j|t_k} \tilde{\mbf{R}}\tilde{\mbf{u}}^\text{AMT}_{j|t_k}\Bigl{)} + (\mbf{x}_{N|t_k}^\text{MPC}-\mbfbar{x}_{N|t_k}^\text{MPC})^\trans \mbf{Q}_N (\mbf{x}_{N|t_k}^\text{MPC}-\mbfbar{x}_{N|t_k}^\text{MPC}) + \mbs{\alpha}^\trans\mbf{C}\mbs{\alpha} \label{eq:MPC}\\
    &\hspace{0em}\text{subject to } \nonumber\\
    &\hspace{1em} \mbf{x}^\text{MPC}_{j+1|t_k} = \mbf{A}^\text{MPC}_{j|t_k}\mbf{x}^\text{MPC}_{j|t_k} + \mbf{B}^\text{MPC}_{w,j|t_k}\mbf{w}_{j|t_k} + \mbf{B}^\text{MPC}_{u1,j|t_k}\mbf{u}^{\text{AMT}}_{j|t_k} + \mbf{B}^\text{MPC}_{u2,j|t_k}{u}^{\text{RCD}}_{j|t_k} + \mbf{z}^\text{MPC}_{j|t_k}, \quad j=0, 1, \ldots, N-1,  \nonumber\\
    &\hspace{1em} \mbf{x}^\text{MPC}_{0|t_k} = \mbf{x}^\text{MPC}(t_k), \nonumber\\
    &\hspace{1em} \mbf{u}^\text{AMT}_{-1|t_k} = \mbf{u}^\text{AMT}(t_k), \nonumber\\
    &\hspace{1em} \mbf{x}^\text{MPC}_{\text{min}} \leq \mbf{x}^\text{MPC}_{j|t_k} \leq \mbf{x}^\text{MPC}_{\text{max}}, \hspace{1em} j=0,\ldots,N, \nonumber\\
    &\hspace{1em} \mbf{u}_{\text{min}} \leq \mbf{u}_{j|t_k} \leq \mbf{u}_{\text{max}}, \hspace{1em} j=0,\ldots,N-1, \nonumber\\
    &\hspace{1em} \dot{\mbf{u}}^\text{AMT}_{\text{min}} \leq \frac{\mbf{u}^\text{AMT}_{j|t_k} - \mbf{u}^\text{AMT}_{j-1|t_k}}{\Delta t} \leq \dot{\mbf{u}}^\text{AMT}_{\text{max}}, \hspace{1em} j=0,\ldots,N-1, \nonumber\\
    &\hspace{1em} \mbf{h}_{4,\text{min}}^\text{soft} - \mbs{\alpha} \leq \mbf{h}^\text{RW}_{4,j|t_k} \leq  \mbf{h}_{4,\text{max}}^\text{soft} + \mbs{\alpha}, \hspace{1em} j=0,\ldots,N, \nonumber\\
    &\hspace{1em} \mbs{\alpha} \geq \mbf{0}, \nonumber
\end{align}
where $ \mbs{\alpha} \in \mathbb{R}^{4}$, $\mbs{\mathcal{X}} = \{\mbf{x}^\text{MPC}_{0|t_k},\mbf{x}^\text{MPC}_{1|t_k},\ldots, \mbf{x}^\text{MPC}_{N|t_k}\}$, $\mbs{\mathcal{U}} = \{\mbf{u}_{-1|t_k},\mbf{u}_{0|t_k},\mbf{u}_{1|t_k},\ldots, \mbf{u}_{N-1|t_k}\}$ are the design variables, $N$ is the number of timesteps in the prediction horizon, $\mbf{x}^\text{MPC}(t_k) = \bbm \hat{\mbs{\theta}}_k^\trans & \hat{\mbs{\omega}}^{ba^\trans}_{b,k} & \mathbf{M}_{34}^\dagger\hat{\mbf{h}}_{b,k}^{\text{RW}^\trans} & \hat{\mbf{e}}_k^{\text{int}^\trans} \ebm^\trans$ is the known system state at time $t_k$, $\mbf{w}_{j|t_k} = \hat{\mbf{w}}^+_{k}$ is the Kalman filter disturbance estimate, and $\mbf{u}^\text{AMT}(t_k)$ is the AMT position at time $t_k$. 

Due to the use of a quadratic objective function, affine equality constraints, and affine inequality constraints, this MPC policy can be solved as a QP at each time step. 
The MATLAB function \texttt{quadprog} with its default settings is used to solve the QP optimization problem in the simulation results of this work.
Solving this problem yields a sequence of optimal control inputs over the prediction horizon, \ie, $\mbs{\mathcal{U}}^* =\{ \mbf{u}_{-1|t_k}^*,\mbf{u}_{0|t_k}^*,\mbf{u}_{1|t_k}^*,\ldots, \mbf{u}_{N-1|t_k}^*\}$. Only the first input ($\mbf{u}_{0|t_k}^*$) is applied to the system before proceeding to the next time step and again solving for the optimal sequence of control inputs.

In this work, the Kalman filter is designed to operate at the same rate (every $100$ seconds) as the MPC momentum management time step.
At every time step $t_k$, a measurement update is performed, and the Kalman filter state $\hat{\mbf{X}}_k^+ = \bbm \hat{\mbf{x}}^+_{k} \\ \hat{\mbf{w}}^+_{k} \ebm$ is extracted to formulate the MPC prediction model in Eq.~\eqref{eq:MPC_pred_model}, where $\mbf{x}_k = \hat{\mbf{x}}^+_{k}$ and $\mbf{w}_k = \hat{\mbf{w}}^+_{k}$ are used for the prediction model at time $t_k$, and the linear transformation $\mbf{T}$ is used to compute $\mbf{x}^\text{MPC}_k$, where $\mbf{x}^\text{MPC}_k = \mbf{T}\mbf{x}_k$.
This transformation enables MPC to seek a minimum norm angular momentum allocation while directly constraining the angular momentum on each RW.

The recursive nature of the MPC necessitates the prediction model to be re-linearized about the current state and inputs at every time step. To improve actuation efficiency and mitigate noise, operational actuation thresholds on the AMT and RCDs are introduced as additional design tuning parameters. 
These thresholds are designed to trim out minor control demands, removing small AMT movements and RCD thrusts that typically arise from minor momentum management or noisy state estimates. 
Specifically, any element of the MPC-demanded AMT position change satisfying the element-wise inequality $\big{|}(\mbf{u}^\text{AMT}_{0|t_k} - \mbf{u}^\text{AMT}_{-1|t_k})/\Delta t\big{|} \leq \beta^\text{AMT}_\text{thresh} \dot{\mbf{u}}^\text{AMT}_{\text{max}}$ is set to stay at its current position ($\mbf{u}^\text{AMT}_{0|t_k} = \mbf{u}^\text{AMT}_{-1|t_k}$) for the upcoming time step. Additionally, if the MPC-demanded RCD input satisfies $\big{|}{u}^{\text{RCD}}_{0|t_k}\big{|} < \beta^\text{RCD}_\text{thresh}{u}^{\text{RCD}}_{\text{max}}$, then it is set to zero.
In both cases, the control input is applied to the system only when the MPC demands an input exceeding the predefined magnitude thresholds.
The momentum management inputs filtered by the thresholds are then passed to perform the time update of the Kalman filter, and applied to the system.

Since the MPC demanded RCD input is a continuous value between $\pm\tau^\text{RCD}_{b3,\text{on}}$, it does not directly match the on-off actuation mechanism of the RCD array.
A single pulse PWM-quantization technique in Eq.~\eqref{eq:PWM_cases} is used to turn the continuous RCD input value to a pulse length specified time with $\tau^\text{RCD}_{b3,\text{on}}$ value.
These thresholding filter and PWM-quantization are leveraging the MPC recursive nature. Once an input is trimmed or modified at one time step, the MPC recalculates the optimal inputs using the latest state at the next time step, compensating for the mismatched input and system dynamics. 


\section{Numerical Simulation Results} \label{sec:NumSim}

Numerical simulation experiments are performed to validate the MPC momentum management policy on Solar Cruiser.
Section~\ref{sec:Sim_Setup} presents the setup of the system and the controller.
Section~\ref{sec:Sim_NASA} presents the state-of-the-art thresholding momentum management control developed for NASA's Solar Cruiser by~\cite{Inness2023MM,Tyler2024}, which is used as a validation of the simulation environment and a benchmark comparison to the proposed method. 
Section~\ref{sec:Sim_KFMPC} presents simulations of the proposed MPC momentum management policy under different conditions, exhibiting the importance of incorporating a disturbance estimate with the MPC policy and the effect that threshold design has on actuator efficiency.
Section~\ref{sec:NumSim_Comparison} presents a direct comparison of the proposed MPC-based momentum management and the state-of-the-art NASA's thresholding method.


\subsection{Simulation Setup} \label{sec:Sim_Setup}

\begin{table}[b!]
\caption{\label{tab:table1} System parameters used in the numerical simulations.}
\centering
\begin{tabular}{cccc}
Parameter & Value & Units \\
\hline
$m_p$ & $55.5$ & kg \\
$m_s$ & $55.5$ & kg \\
$ \mbf{J}^{\mathcal{P}p}_b$ & $\textrm{diag}(4.01, 4.01, 6.07)$ & kg$\cdot$m$^2$\\
$\mbf{J}^{\mathcal{S}s}_b$ & $\textrm{diag}(8049.8, 8049.8, 16099.6)$ & kg$\cdot$m$^2$\\
$r_w$& $0.11$ & m \\
${r}^{ps}_{b3}$ & $0.47$ & m\\
\hline
\end{tabular}
\end{table}   

The simulation parameters are chosen to reflect NASA’s Solar Cruiser~\citep{johnson2019solar, JohnsonLes2020SCTM, Tyler2024, Inness2023MM}. 
The total mass of the sailcraft is $111$~kg, where the bus and the sail each contribute half of the total mass.
The out-of-plane offset between the CM and the sail surface (also the CP) is captured by ${r}^{ps}_{b3}$, the third component of $\mbf{r}^{ps}_b(t)$.
Unlike the simulations performed by~\cite{shen2025} that assumed this offset to be zero, this distance is chosen as ${r}^{ps}_{b3} = 0.47$~m in this paper to more accurately simulate Solar Cruiser's geometry. 
The rest of Solar Cruiser's physical parameters are included in Table~\ref{tab:table1}.

The configuration of the 4-RW pyramid is given by $\psi_i = 60^\circ$ for all $i$, $\phi_i = 45^\circ+(i-1)\cdot90^\circ$ for $i = 1,2,3,4$, resulting in the mapping matrix given in Eq.~\ref{eq:RW_mapping_matrix}. 
The RWs perform attitude tracking using the PID control law in Eq.~\eqref{eq:PID_law} with gains $\mbf{K}_p = 0.25\cdot\mbf{1}_{3\times 3}$~N$\cdot$m/rad, $\mbf{K}_d = 112 \cdot\mbf{1}_{3\times 3}$~N$\cdot$m$\cdot$s/rad, $\mbf{K}_i = 8\times10^{-4}\cdot \mbf{1}_{3\times 3}$~N$\cdot$m/(rad$\cdot$s).
The desired maneuver trajectory ($\mbs{\theta}_d$ and $\mbs{\omega}_d$) follows the sequence of attitude hold, forward slew, attitude hold, return slew, and attitude hold. 
The trajectory parameters include the maximum slew rate $\dot{\mbs{\theta}}_\text{slew} = \big[ 0.01 \,\,\, 0.01 \,\,\, 0.0035\big] ^\trans$~deg/s, acceleration time (required to reach maximum slew rate) $t_\text{accel} = 500$~s, start time of forward slew $t_\text{start} = 10000$~s, and start time of return slew $t_\text{return} = 20000$~s. The desired slew trajectory follows the trapezoidal Euler angle rate profile defined in the Appendix. The initial attitude ${\mbs{\theta}}_0$ and slew goal attitude $\mbs{\theta}_\text{goal}$ are defined in the subsequent simulation cases.

\begin{figure}[b!]
    \centering
        \includegraphics[width=0.99\textwidth]{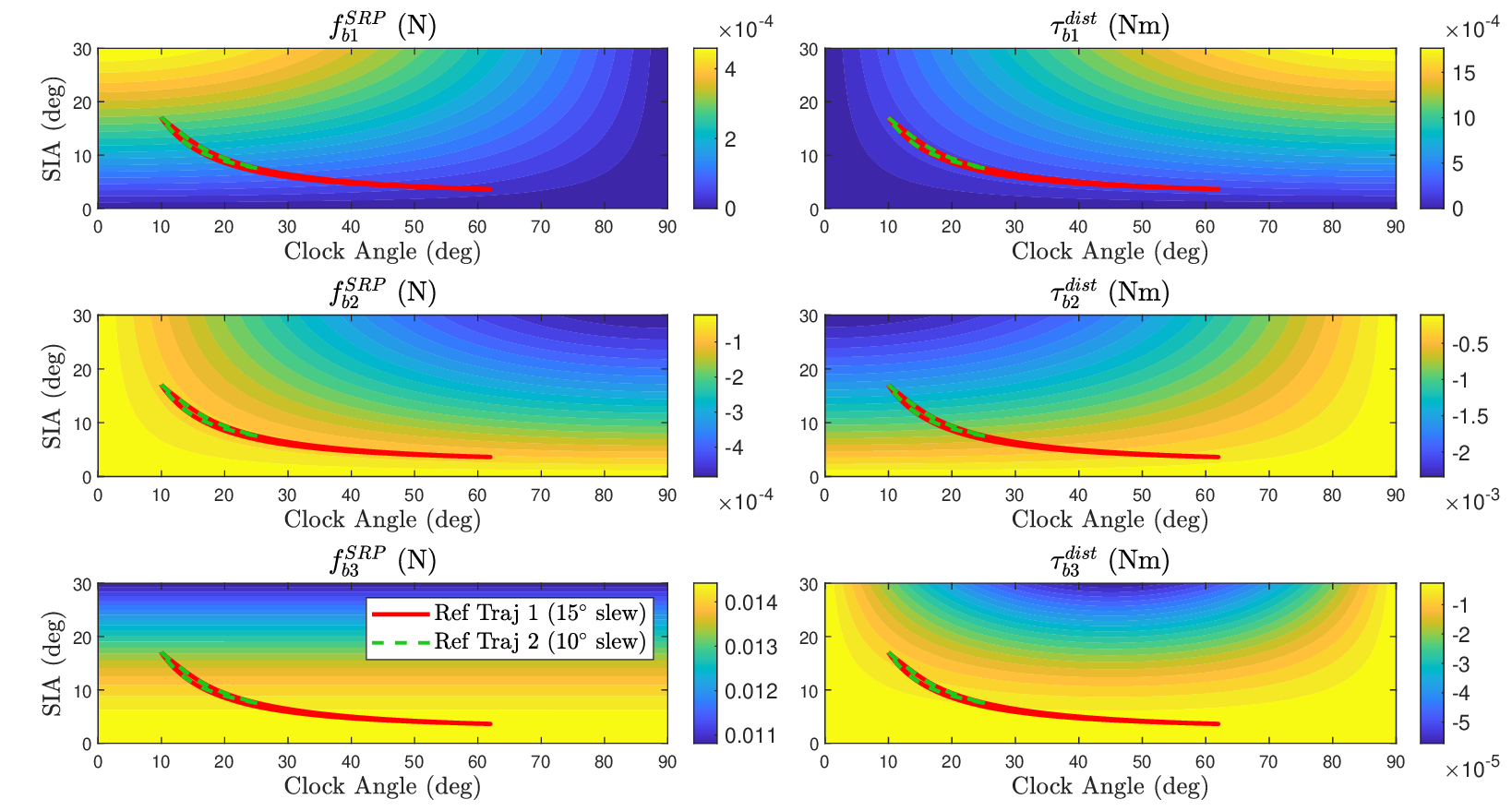}
    \caption{The attitude-dependent SRP force and disturbance torque as functions of SIA and clock angle. The colorbars on the left plots indicate the SRP force, while the colorbars on the right plots indicate the SRP torque. Note that the colorbar scales vary across subplots to accommodate the large magnitude differences between axes. The red lines and the green dashed lines are the reference trajectories defining the operational envelopes for slew maneuvers toward ${\mbs{\theta}}_\text{goal} = \Big[ 0^\circ \,\,\, 15^\circ \,\,\, 1^\circ \Big]^\trans$ and ${\mbs{\theta}}_\text{goal} = \Big[ 0^\circ \,\,\, 10^\circ \,\,\, 1^\circ \Big]^\trans$, respectively, both starting from an initial attitude of $\mbs{\theta}_0 = \mbf{0}$.}\label{Fig:SRP_profile}
\end{figure}

A static membrane shape model developed in~\cite{bunker2026static} and motivated by the Solar Cruiser shape analysis performed by~\cite{gauvain2023solar} is employed to compute $\mbf{f}_b^{\text{SRP}}$ and $\mbs{\tau}_b^{\text{dist}}$ based on the solar sail's attitude $\mbs{\theta}$.
This sail shape model implements Solar Cruiser's membrane optical properties~\citep{heaton2015update,heaton2017near}, as well as its expected structural and membrane deformation~\citep{gauvain2023solar}. A non-flat sail membrane with non-ideal reflectivity properties is used in the numerical simulations, where the billowing at the centroid of each quadrant is $15$~cm, and a $\pm 50$~cm alternating boom tip deflections to match the analysis of~\cite{bunker2026static}. 
The sail membrane shape model from~\cite{bunker2026static} includes a discretized mesh with 3,600 triangular elements, where the local SIA is computed for each planar element. The force exerted on each triangular sail mesh element by the local SRP is applied at the element's planar centroid and modeled as a normal force, $F_n^i$, and tangential force, $F_t^i$,  given by 
\begin{align*}
    F_t^i &= PA^i(1 - rs)\cos(\alpha^i)\sin(\alpha^i), 
    \\
    F_n^i &= -PA^i(1 - rs)\cos^2(\alpha^i) -PA^i B_f(1-s)r\cos(\alpha^i) - PA^i(1-r)\cos(\alpha^i) \left(\frac{e_fB_f-e_bB_b}{e_f+e_b} \right), 
\end{align*}
where $P = 4.5391 \times 10^{-6}$~N/m$^2$ is the solar pressure at one astronomical unit (au), $A^i$ is the area of the i'th sail element, $r = 0.91$ is the reflection coefficient, $s=0.94$ is the fraction of specular reflection coefficient, $\alpha^i$ is the local SIA of the i'th sail element, $B_b = 0.67$ and $B_f = 0.79$ are the back and front non-Lambertian coefficients, and $e_b = 0.27$ and $e_f = 0.025$ are the back and front surface emissivity, respectively~\citep{heaton2015update}. The total SRP force and disturbance torque is computed by adding up the normal and tangential forces across all sail membrane elements and accounting for the location of each element which solving for the SRP disturbance torque. 
The resulting attitude-dependent SRP force and disturbance torque profiles are shown in Fig.~\ref{Fig:SRP_profile}, where the attitude is represented by the conventional SIA and clock angle. The nominal reference attitude trajectory involving a hold-slew-hold-slew-hold maneuver sequence, starting from an initial attitude of $\mbs{\theta}_0 = \mbf{0}$ and slewing to a target attitude of ${\mbs{\theta}}_\text{goal} = \Big[ 0^\circ \,\,\, 15^\circ \,\,\, 1^\circ \Big]^\trans$ is represented by the red lines in Fig.~\ref{Fig:SRP_profile}. The green dashed lines in Fig.~\ref{Fig:SRP_profile} are the nominal reference attitude trajectory with ${\mbs{\theta}}_\text{goal} = \Big[ 0^\circ \,\,\, 10^\circ \,\,\, 1^\circ \Big]^\trans$. The plots are restricted to the $0^\circ$ to $30^\circ$ SIA range, which is twice the nominal operational envelope of the Solar Cruiser mission~\citep{Tyler2024,heaton2023RCD}.
The relationship between sailcraft attitude $\mbs{\theta}$, the associated SIA and clock angle, and the corresponding SRP force and disturbance torque at $1$~au is included in Table~\ref{tab:table2}.
Given the slow varying nature and small magnitude of SRP force and torque, $\mbf{f}^\text{SRP}_b$ and $\mbs{\tau}^\text{dist}_b$ are updated at every $20$ seconds in the simulation.
The simulation and attitude control timestep is $\dee t = 1$~second, and the momentum management time step is $\Delta t = 100$~seconds.

\begin{table}[t!]
\caption{\label{tab:table2} Attitude correspondence to SRP force and torque.}
\centering
\begin{tabular}{cccc}
Attitude $\mbs{\theta}$ & [SIA,clock angle] & $\mbf{f}^\text{SRP}_b$ & $\mbs{\tau}^\text{dist}_b$ \\
\hline
$\mbf{0}$ & [$17^\circ$, $10^\circ$] & $\big[ 3.13 \,\,\, -0.56 \,\,\, 133.03\big] ^\trans\times 10^{-4}$~N & $\big[ 208.78 \,\,\, -1493.80 \,\,\, -6.91\big] ^\trans \times 10^{-6}$~N$\cdot$m \\
$\big[ 0^\circ \,\,\, 10^\circ \,\,\, 1^\circ\big] ^\trans$ & [$7.43^\circ$, $25.57^\circ$] & $\big[ 1.33 \,\,\, -0.64 \,\,\, 143.46\big] ^\trans\times 10^{-4}$~N & $\big[ 238.36 \,\,\, -628.25 \,\,\, -3.11\big] ^\trans \times 10^{-6}$~N$\cdot$m \\
$\big[ 0^\circ \,\,\, 15^\circ \,\,\, 1^\circ\big] ^\trans$ & [$3.62^\circ$, $61.96^\circ$] & $\big[ 0.34 \,\,\, -0.64 \,\,\, 145.39\big] ^\trans\times 10^{-4}$~N & $\big[ 239.98 \,\,\, -161.19 \,\,\, -0.80\big] ^\trans \times 10^{-6}$~N$\cdot$m \\
\hline
\end{tabular}
\end{table}
  
The AMT and RCD actuation is as discussed in Section~\ref{sec:MM_actuation}.
The AMT has a translation limit of $\mbf{u}^\text{AMT}_{\text{max}} = -\mbf{u}^\text{AMT}_{\text{min}} = \Big[ 0.29 \,\,\, 0.29\Big]^\trans$~m and a rate limit of $\dot{\mbf{u}}^\text{AMT}_{\text{max}} = -\dot{\mbf{u}}^\text{AMT}_{\text{min}} = \Big[ 0.5 \,\,\, 0.5\Big]^\trans$~mm/s in the $\vect{b}^1$ and $\vect{b}^2$ axes~\citep{JohnsonLes2020SCTM}.
The discrete-time rate constraint is defined as $\dot{\mbf{u}}^\text{AMT}_{\text{max}} = -\dot{\mbf{u}}^\text{AMT}_{\text{min}} = (\mbf{u}^\text{AMT}_{j+1|t_k} - \mbf{u}^\text{AMT}_{j|t_k})/\Delta t$, which limits the maximum AMT position change to be $0.05$~m in each axis at every momentum management time step. In the simulation ($\dee t = 1$~second), the AMT is actuated to move at its maximum rate towards the momentum management commanded position (updated every $\Delta t = 100$~seconds) in each of its axis until reaching the commanded position. The roll torque generated when the RCDs are turned on is set to meet the Solar Cruiser’s roll torque requirement as defined in Eq.~\eqref{eq:RCD_torque}.

The Kalman filter measurement noise covariance is chosen based on NASA Solar Cruiser's performance requirement~\citep{JohnsonLes2020SCTM}.
It is assumed that the onboard ADCS measurement noise standard deviation is 3 times smaller (more accurate) than the standard deviation of control requirement defined by~\cite{JohnsonLes2020SCTM}.
Based on the required pointing accuracy of $<60$~arcsec in pitch/yaw and $<6.8$~arcmin in roll ($3\sigma$), the standard deviation of attitude measurement noise is chosen as $0.00186^\circ$ in pitch/yaw and $0.0125^\circ$ in roll, \ie, $\mbs{\sigma}_\theta = \text{diag}(0.00186^\circ, 0.00186^\circ, 0.0125^\circ)$.
Based on the pointing jitter requirements of $<10$~arcsec/sec in pitch/yaw and $<1.34$~arcmin/sec in roll ($3\sigma$), the standard deviation of angular rate measurement noise is chosen as $0.000306$~deg/sec in pitch/yaw and $0.0025$~deg/sec in roll, \ie, $\mbs{\sigma}_\omega = \text{diag}(0.000306, 0.000306, 0.0025)$~deg/sec.
It is assumed that the RW angular momentum measurement accuracy is $\mbs{\sigma}_h = 10^{-5} \cdot \mbf{1}_{3\times3}$~N$\cdot$m$\cdot$s, and $\mbs{\sigma}_e = 10^{-8} \cdot \mbf{1}_{3\times3}$~rad$\cdot$s, as these parameters are not publicly available in the literature.
The collective measurement noise covariance is given by $\mbf{R}^\text{KF} = \text{diag}(\mbf{r}_\theta, \mbf{r}_\omega, \mbf{r}_h, \mbf{r}_e) = \text{diag}(\mbs{\sigma}_\theta^2, \mbs{\sigma}_\omega^2, \mbs{\sigma}_h^2, \mbs{\sigma}_e^2)$.
In the simulation, zero-mean Gaussian white noise with the same measurement covariance is added to each of the measurement parameters in Kalman filter measurement update step.

The Kalman filter process noise covariance is given by $\mbf{Q}^\text{KF} = \text{diag}(\mbf{Q}^\text{KF}_\text{model}, \mbf{Q}^\text{KF}_\text{dist})$, which is largely a tuning parameter of the filter.
The dynamic model error covariance $\mbf{Q}^\text{KF}_\text{model} = \text{diag}(0.01^2\cdot \mbf{1}_{3\times3}, 0.0001^2\cdot \mbf{1}_{3\times3}, 10^{-6}\cdot \mbf{1}_{3\times3}, 10^{-16}\cdot \mbf{1}_{3\times3})$, whose units are deg$^2$,deg$^2/$s$^2$,(N$\cdot$m$\cdot$s)$^2$, and (rad$\cdot$s)$^2$, characterizes the combination of linearization error in the dynamics and expected deviations in the trajectory.
The disturbance model error covariance $\mbf{Q}^\text{KF}_\text{dist} = \text{diag}( {q}_{\tau, 1}+\xi_1\omega_{d,1}^2, {q}_{\tau,2}+\xi_2\omega_{d,2}^2, {q}_{\tau,3} + \xi_3 \omega_{d,3}^2)$ has static covariance ${q}_{\tau, 1} = {q}_{\tau, 2} = 5\times 10^{-6}$, ${q}_{\tau, 3} = 5\times 10^{-9}$, and dynamic covariance scaling parameters $\xi_1 = \xi_2 = 0.5$, $\xi_3 = 0.1$. The dynamic covariance parameters account for the fact that the attitude, and thus, the disturbance torques, are expected to deviate more quickly when performing a slew maneuver. The resulting covariance  $\mbf{Q}^\text{KF}_\text{dist} = \text{diag}( 5\times 10^{-6}+\onehalf\omega_{d,1}^2, 5\times 10^{-6}+\onehalf\omega_{d,2}^2, 5\times 10^{-9} + 0.1 \omega_{d,3}^2)$~N$\cdot$m$^2$ characterizes the slowly-varying nature of the disturbance estimate and the other model discrepancies captured by $\hat{\mbf{w}}$.
The initial state estimate $\hat{\mbf{X}}^-_0 = \mbf{0}$ does not consider any preliminary information of the state and disturbance error.
The initial estimation error covariance is chosen as $\mbf{P}_0^- = \text{diag}(100\mbf{Q}^\text{KF}_\text{model}, \text{diag}(10^{3},10^{3},10^{9})\mbf{Q}^\text{KF}_\text{dist})$ to allow initial estimate correction. 

In this work, the Kalman filter operates at the same frequency as the momentum management system, which has a time step of $100$ seconds.
The system undergoes an initial slew of attitude tracking, and the Kalman filter acquires its first measurement update at the first momentum management timestep, \ie, $t_k = 100$~sec.
After the measurement update, the momentum management policy determines the associated AMT and RCD inputs, which are then used in the time update using the Kalman filter process model.
The momentum management input commands are passed through the AMT and RCD actuation dynamics as discussed in Section~\ref{sec:MM_actuation}, and then applied to the nonlinear dynamics as in Eq.~\eqref{EOM} until the next momentum management timestep. The process of a measurement update, momentum management input determination, time update, and application of the input to the nonlinear system is repeated.


\subsection{NASA's State-of-the-art Method} \label{sec:Sim_NASA}
NASA's state-of-the-art momentum management strategy used on Solar Cruiser is establishes as a benchmark comparison to the proposed MPC strategy.
The Solar Cruiser momentum management system utilizes three threshold-based decoupled channels to command the AMT and RCDs~\citep{Inness2023MM,Tyler2024}.
Solar Cruiser employs on-off thresholds for both AMT and RCD activation, which are based on the RWs' stored angular momentum in the pitch/yaw and roll axes. An upper activation threshold is set higher than a lower deactivation threshold, establishing a hysteresis. 
Specifically, an actuator engages only when its corresponding RW momentum exceeds the activation threshold and remains active until the momentum drops below the deactivation threshold. 

The two AMT axes (pitch and yaw) are controlled independently via PID control laws, which regulate the accumulated angular momentum stored in the corresponding RWs. The control laws for the two axes are defined as
\begin{align*}
    u^{\text{AMT}}_{1} &= K^\text{AMT}_p h^\text{RW}_{b2} +K^\text{AMT}_d \dot{h}^\text{RW}_{b2} +K^\text{AMT}_i \int^t_{t_0} h^\text{RW}_{b2}(\tau) \dee\tau,\\
    u^{\text{AMT}}_{2} &= -K^\text{AMT}_p h^\text{RW}_{b1} -K^\text{AMT}_d \dot{h}^\text{RW}_{b1} -K^\text{AMT}_i \int^t_{t_0} h^\text{RW}_{b1}(\tau) \dee\tau.
\end{align*}
The sign difference between the two PID control laws reflects the dynamics in Eq.~\eqref{EOM}, where the AMT-induced torque $\mbs{\tau}_b^{\text{AMT}} = \f{m_s}{m_p+m_s}\mbf{r}_b^{ps^\times} \mbf{f}_b^{\text{SRP}}$ involves a cross product with opposite signs along the body 1 and 2 axes.
This control input is updated with a time step of $\Delta t = 100$~sec using a ZOH to maintain a constant command throughout the interval.
The RCDs' actuation follows a simple on-off logic with a fixed torque magnitude when activated. The RCD activation/deactivation switch aligns with the momentum management time step $\Delta t$.

This threshold-based control, along with the AMT PID gains, is tuned via simulation to optimize performance.
Crucially, this PID control framework does not inherently account for physical actuator constraints, such as AMT position and rate limits. These limits are enforced externally after the PID controller determines the position command. Consequently, tuning the controller to ensure effective momentum management while avoiding actuator saturation remains a key design challenge.

In the absence of any numerical values in the work of~\cite{Inness2023MM,Tyler2024}, values are chosen in this paper in an attempt to recreate the results of~\cite{Inness2023MM,Tyler2024}. To this end, the chosen thresholds for the AMT are $0.25$~N$\cdot$m$\cdot$s for activation, and $0.125$~N$\cdot$m$\cdot$s for deactivation.
The PID gains of the AMT controller are chosen as $K^\text{AMT}_p = 0.1$~(N$\cdot$s)$^{-1}$, $K^\text{AMT}_d = 0.05$~N$^{-1}$, and $K^\text{AMT}_i = 0.0001$~N$^{-1}$s$^{-2}$.
The maximum position constraint of the AMT is enforced such that $|u^{\text{AMT}}_{i}| = u^{\text{AMT}}_{i,\text{max}} = 0.29$~m when the determined PID controller input satisfies $|u^{\text{AMT}}_{i}| > u^{\text{AMT}}_{i,\text{max}}$ ($i = 1, 2$).
The maximum AMT rate constraint is enforced such that $|u^{\text{AMT}}_{i}| = \Delta t \cdot \dot{u}^{\text{AMT}}_{i,\text{max}} = 0.05$~m when the determined PID input satisfies $|u^{\text{AMT}}_{i}| > \Delta t \cdot \dot{u}^{\text{AMT}}_{i,\text{max}}$ ($i = 1, 2$).
The chosen RCD thresholds are $0.125$~N$\cdot$m$\cdot$s for activation, and $0.312$~N$\cdot$m$\cdot$s for deactivation.

For practicality and for a fair comparison to the proposed method, the threshold-based momentum management uses state estimates from the Kalman filter to determine AMT and RCD inputs. Specifically, the angular momentum estimate $\hat{\mbf{h}}_{b,k}^{\text{RW}^+}$ is used to assess the activation/deactivation threshold and AMT proportional control, and $\hat{\dot{\mbf{h}}}_{b,k}^{\text{RW}^+} = \mbf{K}_p(\hat{\mbs{\theta}}^+_k-\mbs{\theta}_d)+\mbf{K}_d(\hat{\mbs{\omega}}^{{ba}^+}_{b,k}-\mbs{\omega}_d)+\mbf{K}_i\hat{\mbf{e}}^{\text{int}^+}_k$ is used for the AMT derivative control. For the AMT integral control, it is assumed that a perfect measurement of $\int^t_{t_0} \mbf{h}^\text{RW}_{b}(\tau)\dee\tau$ is accessible in the ADCS.

\begin{figure}[t!] 
\centering
\subfigure[attitude]
{
        \includegraphics[width=0.48\textwidth]{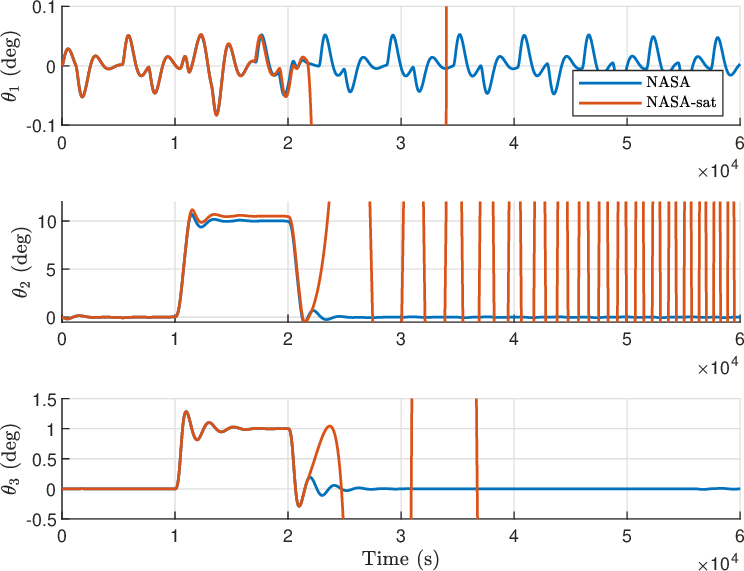}
        \label{Fig_NASA_sima}
}
\subfigure[4 RWs angular momentum]
{
        \includegraphics[width=0.48\textwidth]{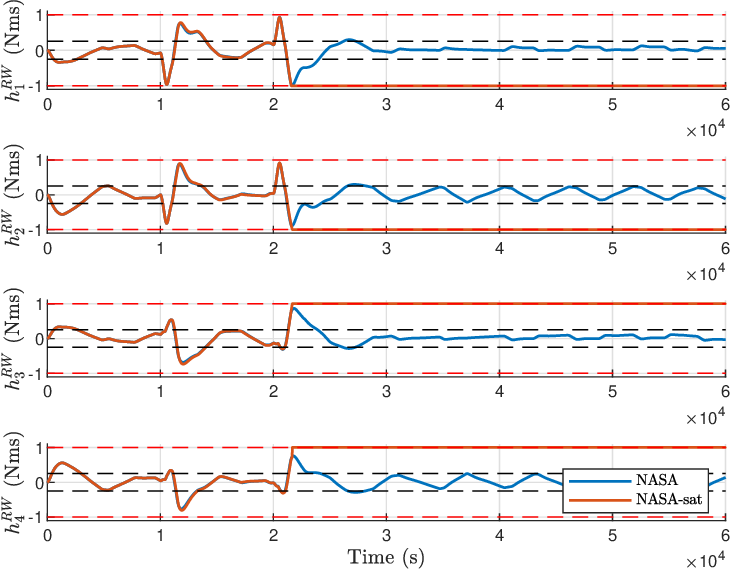}
        \label{Fig_NASA_simb}
}
\\
\centering
\subfigure[body-frame RWs angular momentum]
{
        \includegraphics[width=0.48\textwidth]{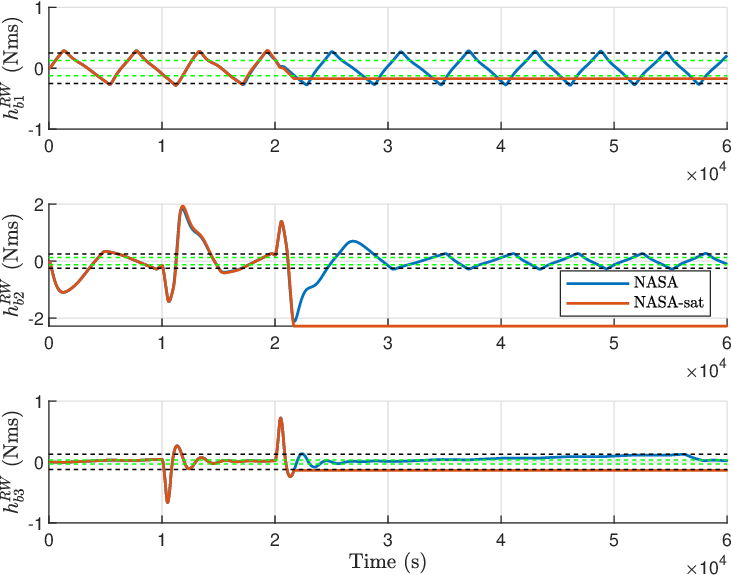}
        \label{Fig_NASA_simc}
}
\subfigure[momentum management inputs]
{
        \includegraphics[width=0.48\textwidth]{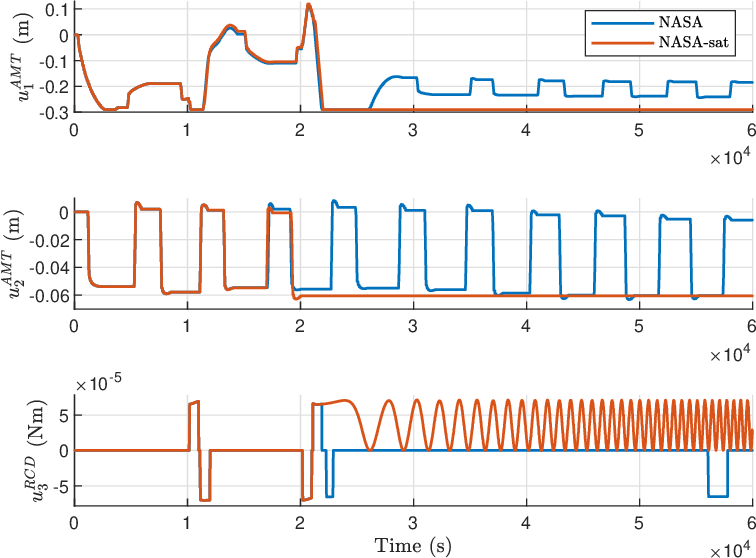}
        \label{Fig_NASA_simd}
}
\centering
\caption{Simulation results using NASA's Solar Cruiser momentum management strategy from~\cite{Inness2023MM,Tyler2024}, featuring RW saturation under a slew of ${\mbs{\theta}}_\text{goal} = \Big[ 0^\circ \,\,\, 10^\circ \,\,\, 1^\circ \Big]^\trans$ trajectory compared to a ${\mbs{\theta}}_\text{goal} = \Big[ 0^\circ \,\,\, 10.5^\circ \,\,\, 1^\circ \Big]^\trans$ trajectory. The black dashed line in (c) denotes the activation threshold ($50\%$ of the soft constraint in MPC) , and the green dashed line denotes the deactivation threshold ($50\%$ of the activation threshold).
}
\label{Fig_NASA_sim}
\end{figure}

Using NASA's state-of-the-art thresholding momentum management policy, a maneuver sequence tracking initial attitude of $\mbs{\theta}_0 = \mbf{0}$ to ${\mbs{\theta}}_\text{goal} = \Big[ 0^\circ \,\,\, 10^\circ \,\,\, 1^\circ \Big]^\trans$ is performed, following the desired hold-slew-hold-slew-hold trajectory.
As shown in Fig.~\ref{Fig_NASA_sim} with the label ``NASA,'' the maneuver demonstrates effective momentum management, where the steady-state performance is similar to that shown by~\cite{Inness2023MM,Tyler2024}, although this is difficult to compare quantitatively due to redacted plot axes.
The momentum management method developed by~\cite{Inness2023MM,Tyler2024} is effective at keeping the angular momentum of the RWs within reasonable bounds with realistic actuation inputs.
However, with a slightly larger slew maneuver with ${\mbs{\theta}}_\text{goal} = \Big[ 0^\circ \,\,\, 10.5^\circ \,\,\, 1^\circ \Big]^\trans$, the system suffers from RW saturation, and the solar sail loses attitude control authority, which is shown in the result of Fig.~\ref{Fig_NASA_sim} with the label ``NASA-sat.'' 
For reference, the black dashed lines in Fig.~\ref{Fig_NASA_simb} indicate $25\%$ of the angular momentum capacity of each RW, which is also the soft constraint value chosen for the proposed MPC-based approach in the following sections.
The black dashed lines in Fig.~\ref{Fig_NASA_simc} indicate the activation thresholds, while the green dashed lines represent the deactivation thresholds.


\subsection{Proposed MPC-based Momentum Management Supported by KF Disturbance Estimate} \label{sec:Sim_KFMPC}

While Solar Cruiser's momentum management method failed to desaturate the RWs and eventually lost attitude control when performing the larger slew, the proposed MPC-based momentum management strategy has the potential to foresee the upcoming angular momentum growth and proactively take actions. This allows for more aggressive slews while maintaining RW control authority.

To highlight this improved performance, simulations of a larger slew maneuver are performed, regulating the reference trajectory of $\mbs{\theta}_0 = \mbf{0}$ and ${\mbs{\theta}}_\text{goal} = \Big[ 0^\circ \,\,\, 15^\circ \,\,\, 1^\circ \Big]^\trans$ with the RW PID control law, while the stored RW angular momentum is unloaded by the momentum management MPC policy outlined in Section~\ref{sec:MPC}. The system parameters and Kalman filter parameters are the same as presented in Section~\ref{sec:Sim_Setup}.
The MPC prediction horizon is chosen as $N = 10$ timesteps, corresponding to a $1000$~sec forecast. 
The state constraints in MPC are determined by mission requirements and RW limits, with the reference attitude tracking limit $\mbs{\theta}_{\text{err}} =\Big[ 5^\circ\,\,\,5^\circ\,\,\,5^\circ\Big]^\trans$, the angular velocity tracking limit $\mbs{\omega}_\text{err} =\Big[ 0.1 \,\,\, 0.1 \,\,\, 0.1 \Big]^\trans$~deg/s, the RW angular momentum capacity $\mbf{h}^{\text{RW}}_{4,\text{max}} = -\mbf{h}^{\text{RW}}_{4,\text{min}} = \Big[ 1 \,\,\, 1  \,\,\,1  \,\,\,1\Big] ^\trans$~N$\cdot$m$\cdot$s, and a large PID integral term $\mbf{e}^{\text{int}}_{\text{max}} = -\mbf{e}^{\text{int}}_{\text{min}} = \Big[ 10^6 \,\,\, 10^6  \,\,\, 10^6\Big] ^\trans$~rad$\cdot$s as an internal state limit.
The attitude and angular rate constraints are set to arbitrarily large limits for design completeness and flexibility, ensuring the framework can accommodate future mission requirements that may involve more aggressive maneuvers.
The soft constraint limits are chosen as $25\%$ of the RWs angular momentum capacity, \ie, $\mbf{h}_{4,\text{max}}^\text{soft} = 0.25\cdot \mbf{h}^{\text{RW}}_{4,\text{max}}$ and $\mbf{h}_{4,\text{min}}^\text{soft} = -\mbf{h}_{4,\text{max}}^\text{soft}$.
The slack variable $\mbs{\alpha} \geq \mbf{0}$ is penalized heavily by the weighting matrix $\mbf{C} = 10000\cdot\mbf{1}_{4\times4}$ in the objective function when $\mbf{h}^\text{RW}_{4,j|t_k}$ deviates from the soft constraint envelope.
The weights in the MPC objective function are provided in Table~\ref{tab:table4}, which are parameters that can be tuned to tailor the performance objective  to different mission stages and scenarios.

\begin{table}[b!]
\caption{\label{tab:table4} MPC tuning parameters used in the numerical simulations.}
\centering
\begin{tabular}{ccc}
Parameter & Value \\
\hline
$N$ & $10$ \\
$\mbf{Q}$ & $\text{diag}(10\cdot\mbf{1}_{6\times6}, 0.5\cdot\mbf{1}_{4\times4}, \mbf{0}_{3\times3})$ \\
$\mbf{Q}_N$ & $10 \cdot \mbf{Q}$ \\
$\mbf{R}$ & $\text{diag}(1,1,5\times10^6)$\\
$\tilde{\mbf{R}}$ & $2000\cdot\mbf{1}_{2\times2}$\\
$\mbf{C}$ & $10000\cdot\mbf{1}_{4\times4}$\\
\hline
\end{tabular}
\end{table} 
It is worth noting that the MPC evaluates RCD inputs as continuous values between $\pm \tau^\text{RCD}_{b3,\text{on}}$, but the actual applied input is quantized into the full on/off value with pulse length $t_c$ using PWM quantization as in Eq.~\eqref{eq:PWM_cases}. 
The current SRP force $\mbf{f}_b^{\text{SRP}}(t_k)$ and RCD on torque $\tau^\text{RCD}_{b3,\text{on}}(t_k)$ used in MPC are updated at every momentum management timestep $t_k$, while the Kalman filter provides the state and disturbance estimates. MPC's assumption that these values are constant across the prediction horizon further shows the robustness of the proposed method.
The relaxation of the on-off RCD actuation constraints allows for the use of off-the-shelf QP solvers that can solve the optimization problem efficiently. The MATLAB function \texttt{quadprog} with its default settings is used. The mean QP solution time across $600$ momentum management timesteps is $32.86$~ms, and the numerical integration time for the $N=10$ LTV prediction model is $39.80$~ms, representing the average of five simulation sets of ${\mbs{\theta}}_\text{goal} = \Big[ 0^\circ \,\,\, 15^\circ \,\,\, 1^\circ \Big]^\trans$ maneuver. For reference, these computations are performed on a desktop computer with a 13th Gen Intel Core i5-13400 @ 2.5~GHz with 24~GB memory and the code is run in Matlab 2024b.

\begin{figure}[t!] 
\centering
\subfigure[attitude]
{
        \includegraphics[width=0.48\textwidth]{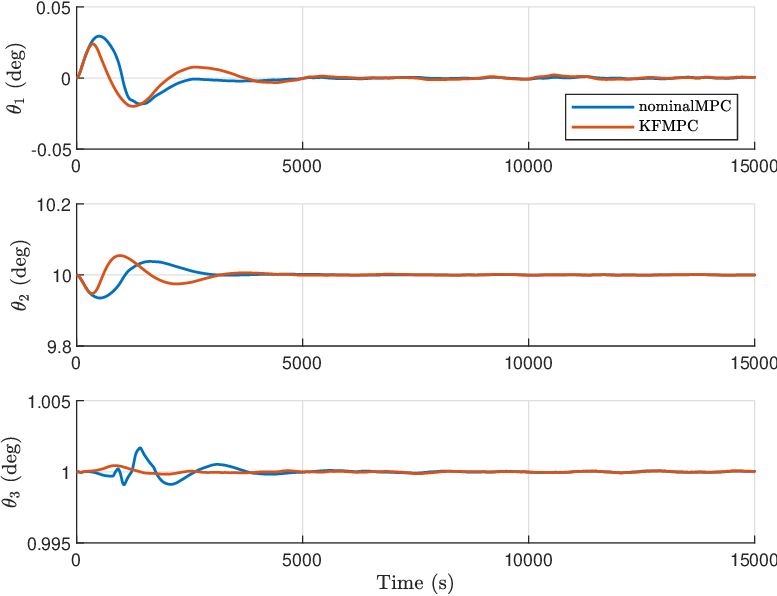}
        \label{Fig_MPC_sima}
}
\subfigure[4 RWs angular momentum]
{
        \includegraphics[width=0.48\textwidth]{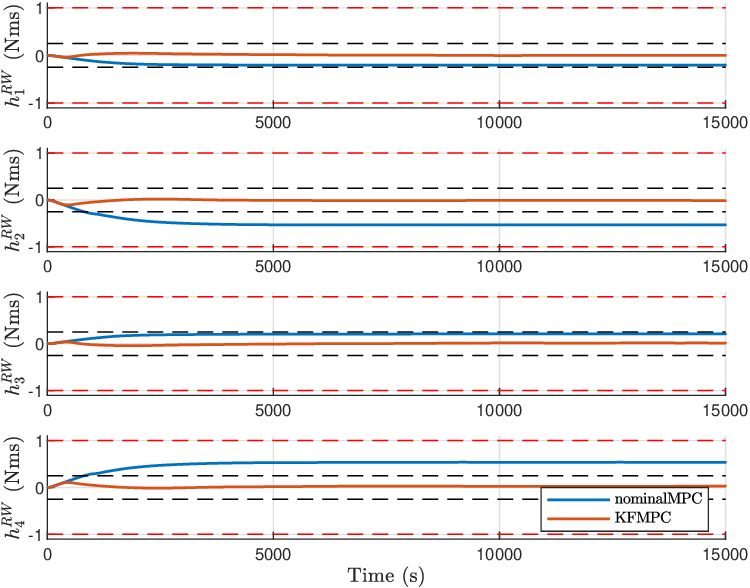}
        \label{Fig_MPC_simb}
}
\\
\centering
\subfigure[body-frame RWs angular momentum]
{
        \includegraphics[width=0.48\textwidth]{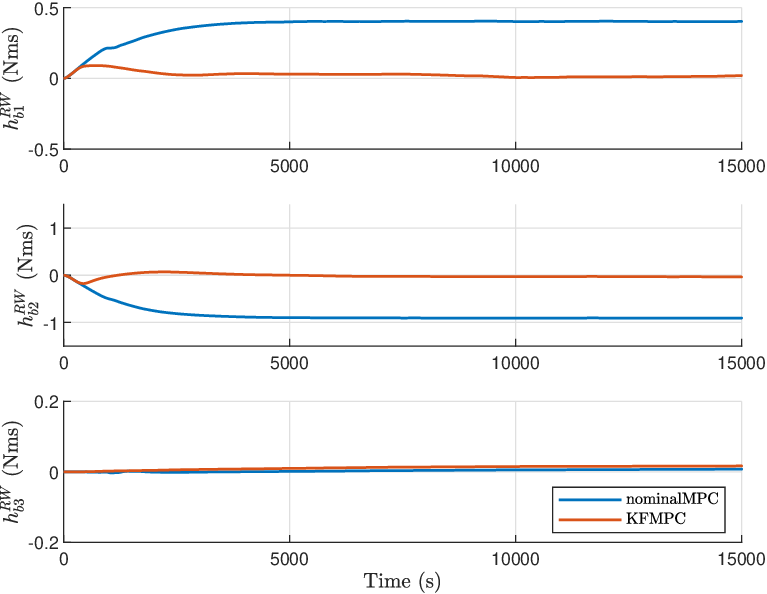}
        \label{Fig_MPC_simc}
}
\subfigure[momentum management inputs]
{
        \includegraphics[width=0.48\textwidth]{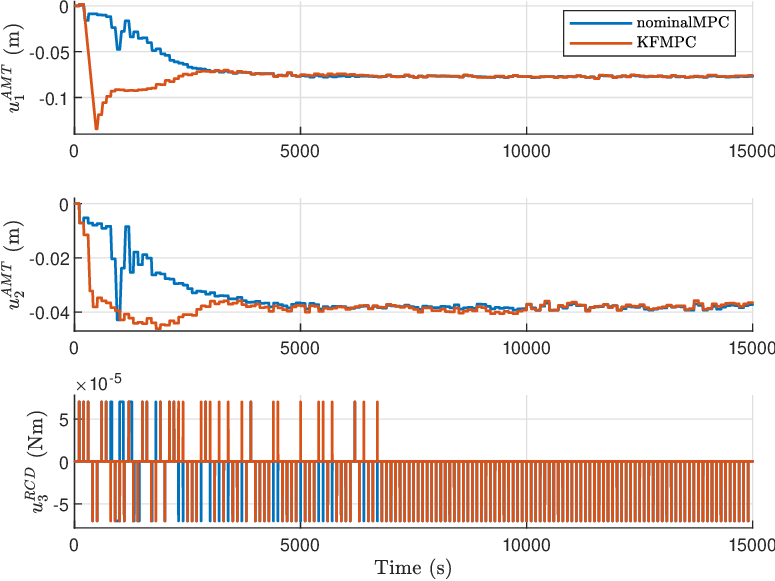}
        \label{Fig_MPC_simd}
}
\centering
\caption{Simulation results using the proposed MPC momentum management strategy under an attitude hold of $\mbs{\theta}_0 = {\mbs{\theta}}_\text{goal} = \Big[ 0^\circ \,\,\, 10^\circ \,\,\, 1^\circ \Big]^\trans$ with and without (nominal) the disturbance estimate knowledge in prediction model. The black dashed lines in (b) denote the $25\%$ soft constraint on 4-RWs angular momentum.}
\label{Fig_MPC_sim}
\end{figure}

To demonstrate the importance of disturbance knowledge in the MPC framework, simulation results with and without the Kalman filter disturbance estimate knowledge in MPC are shown in Fig.~\ref{Fig_MPC_sim}, where no threshold is used ($\beta^\text{AMT}_\text{thresh} = \beta^\text{RCD}_\text{thresh} = 0$). 
The result in blue labeled ``nominalMPC'' uses the nominal MPC implementation without disturbance knowledge, where the MPC prediction model uses $\mbf{w}_{j|t_k} = \mbf{0}$, for $j=0, 1, \ldots, N-1$. The result in red labeled ``KFMPC'' includes the Kalman filter estimate disturbance within the MPC prediction model, where $\mbf{w}_{j|t_k} = \hat{\mbf{w}}^+_{k}$, for $j=0, 1, \ldots, N-1$.
Although both of the MPC policies perform successful momentum management under an attitude hold at $\mbs{\theta}_0 = {\mbs{\theta}}_\text{goal} = \Big[ 0^\circ \,\,\, 10^\circ \,\,\, 1^\circ \Big]^\trans$, the nominal MPC results in the angular momentum of two RWs stabilizing near their saturation limits.
Conversely, the MPC implementation incorporating the disturbance estimate exhibits a significant performance improvement, driving all RW angular momentum down to values safely within the specified soft constraint boundaries, thereby reserving greater control authority. An attitude hold at $\mbs{\theta} = \mbf{0}$ (which has a  higher disturbance torque) and other slew maneuvers have been tested without the disturbance estimate, all of which resulted in RW saturation and instability.
This further shows that the disturbance estimate is critical to the performance of MPC-based momentum management.

\begin{figure}[t!] 
\centering
\subfigure[body-frame RWs angular momentum]
{
        \includegraphics[width=0.48\textwidth]{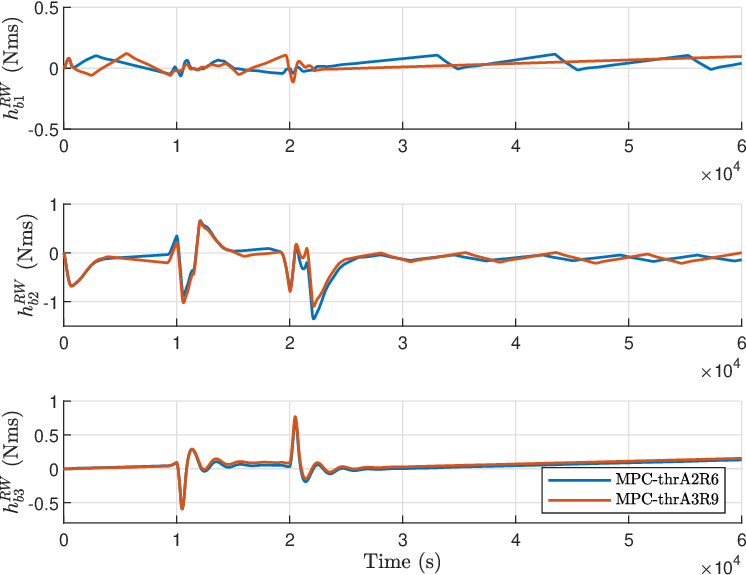}
        \label{Fig_ThrMPC_sima}
}
\subfigure[4 RWs angular momentum]
{
        \includegraphics[width=0.48\textwidth]{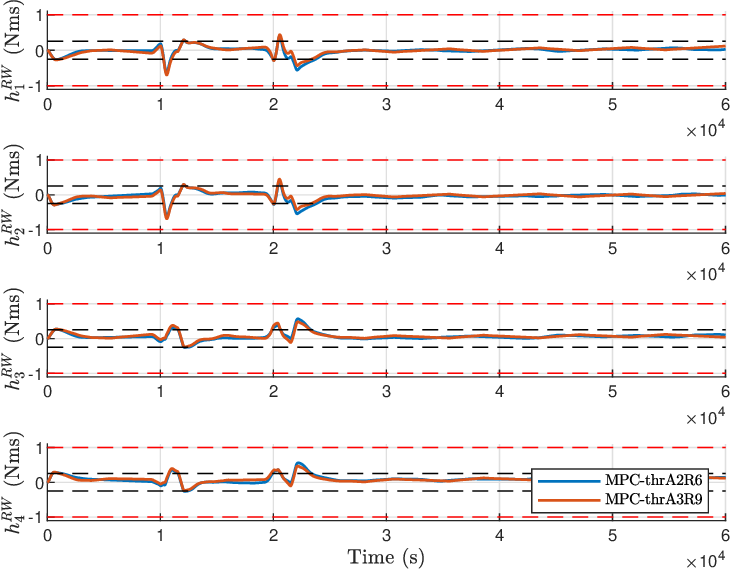}
        \label{Fig_ThrMPC_simb}
}
\\
\centering
\subfigure[momentum management inputs]
{
        \includegraphics[width=0.48\textwidth]{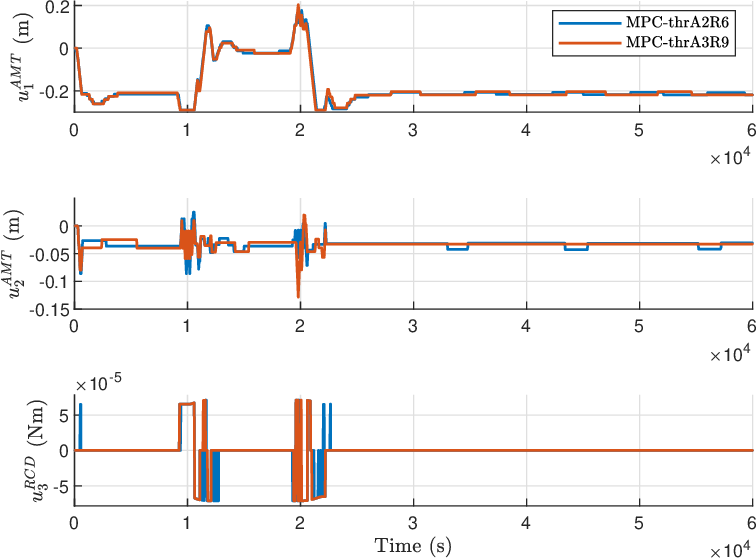}
        \label{Fig_ThrMPC_simc}
}
\subfigure[momentum management inputs (zoomed in)]
{
        \includegraphics[width=0.48\textwidth]{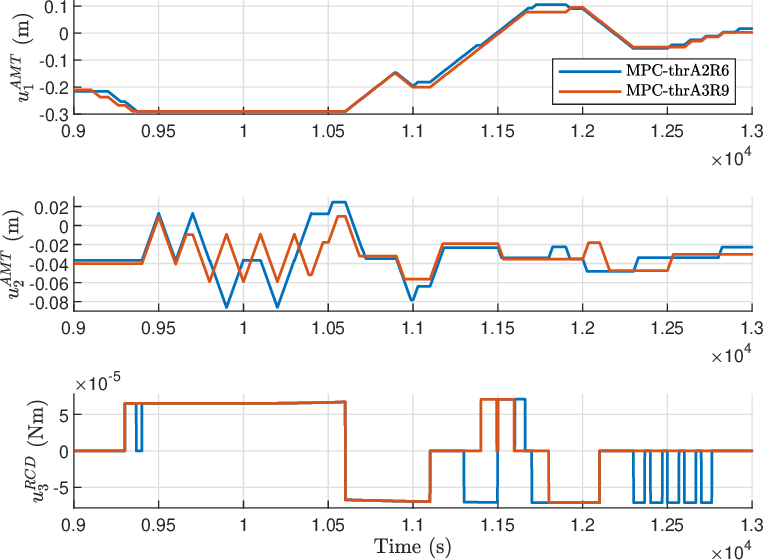}
        \label{Fig_ThrMPC_simd}
}
\centering
\caption{Threshold tuning of the proposed MPC framework under a maneuver sequence tracking initial attitude $\mbs{\theta}_0 = \mbf{0}$ to ${\mbs{\theta}}_\text{goal} = \Big[ 0^\circ \,\,\, 15^\circ \,\,\, 1^\circ \Big]^\trans$ and return, using $20\%$ AMT threshold and $60\%$ RCD threshold (blue), and using $30\%$ AMT threshold and $90\%$ RCD threshold (red). The zoomed-in plot in (d) demonstrates the control inputs between $9000$ to $13000$ seconds of forward slew maneuver and the PWM-quantized RCD actuation pulsing at every time step.}
\label{Fig_ThrMPC_sim}
\end{figure}

The proposed MPC policy with Kalman filter estimation framework is used to perform the maneuver sequence tracking initial attitude of $\mbs{\theta}_0 = \mbf{0}$ to ${\mbs{\theta}}_\text{goal} = \Big[ 0^\circ \,\,\, 15^\circ \,\,\, 1^\circ \Big]^\trans$, following the desired hold-slew-hold-slew-hold trajectory. An additional actuation threshold can be applied on the MPC inputs to filter out minor actuation with minimal loss in momentum management performance.
Leveraging the recursive nature of the MPC, an input activation threshold is applied to trim out minor actuation demanded by MPC, and further improve actuator efficiency and mitigate noise.
A set of results are presented in Fig.~\ref{Fig_ThrMPC_sim}, demonstrating the design choice of actuation thresholds.
The result in blue (labeled ``MPC-thrA2R6'') uses a $20\%$ AMT threshold ($\beta^\text{AMT}_\text{thresh} = 0.2$) and $60\%$ RCD threshold ($\beta^\text{RCD}_\text{thresh} = 0.6$), which means that when MPC demands an AMT input less than $20\%$ of the distance the AMT can move in one direction in one time step ($20\%$ of $0.05$~m), the AMT is held at its current position for the next time step, and the RCD input is set to zero when the MPC-demanded input is less than $60\%$ of the RCD ``on'' torque value.
The result in red (labeled ``MPC-thrA3R9'') uses a $30\%$ AMT threshold ($\beta^\text{AMT}_\text{thresh} = 0.3$) and $90\%$ RCD threshold ($\beta^\text{RCD}_\text{thresh} = 0.9$) on the MPC-demanded inputs.
Figures~\ref{Fig_ThrMPC_sima} and~\ref{Fig_ThrMPC_simb} show that the design choice of the applied thresholds do not degrade momentum management performance, which is further illustrated in the plot of the control inputs in Fig.~\ref{Fig_ThrMPC_simc} and the zoomed in control input plot of Fig.~\ref{Fig_ThrMPC_simd}. 
A comparison of actuation usage among the the three MPC policies with different actuation threshold performing the $15^\circ$ slew maneuver sequence over $60000$ seconds is included in Table~\ref{tab:MPC_CtrlUsage}.
The performance metric of control actuation effort is evaluated by the number of RCD on-off cycles, the total time the RCDs are turned ``on'', the total AMT travel distance in each translation axis, and the sum of the total AMT travel distance across both axes. The design choice of thresholds $\beta^\text{AMT}_\text{thresh}$ and $\beta^\text{RCD}_\text{thresh}$ can be determined by the operational characteristics and the expected lifetime of the actuators.

\begin{table}[b!]
\caption{Momentum management control actuation usage of the proposed MPC policy under different actuation threshold tuning.}
\label{tab:MPC_CtrlUsage}
\centering
\begin{tabular}{c|ccc}
    AMT/RCD Threshold & AMT 10\% / RCD 30\% & AMT 20\% / RCD 60\% & AMT 30\% / RCD 90\% \\
    \hline
    RCD Cycle ($\#$)& 42 & 28 & 12  \\
    RCD On Time (sec)& 6316 & 5554 & 4989 \\
    AMT Dist 1 (cm)& 220.52 & 235.25 & 241.13 \\
    AMT Dist 2 (cm)& 152.24 & 147.82 & 144.74 \\
    Sum of AMT Dist (cm)& 372.76& 383.07& 385.87\\
\end{tabular} 
\end{table}

\begin{figure}[t!]
    \centering
        \includegraphics[width=0.7\textwidth]{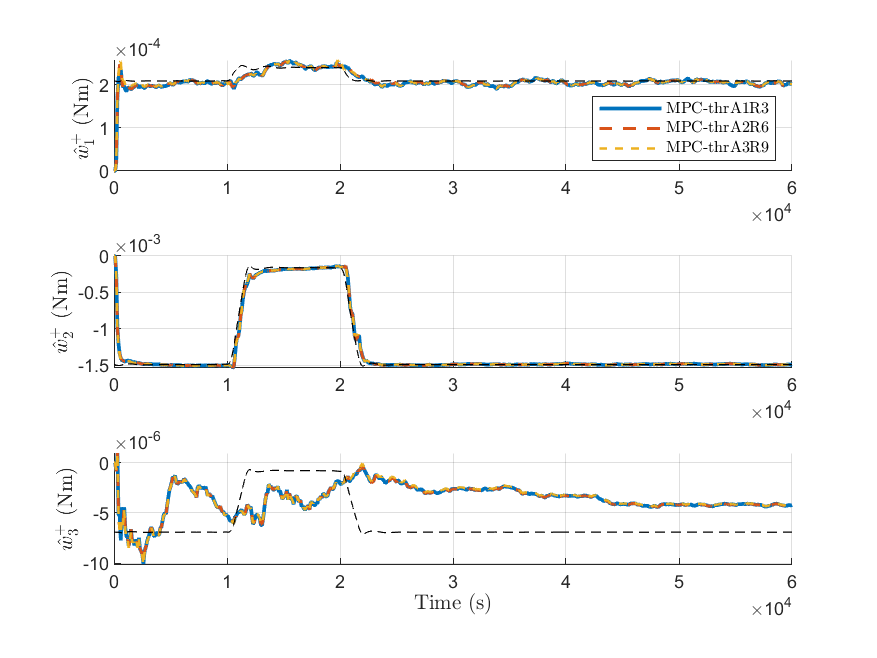}
    \vspace{-8pt}
    \caption{Kalman filter disturbance estimate values used in the MPC momentum management with $10\%$ AMT and $30\%$ RCD threshold (blue), $20\%$ AMT and $60\%$ RCD threshold (red), and $30\%$ AMT and $90\%$ RCD threshold (yellow). The black dashed line indicates the true SRP disturbance torque.}\label{Fig:Dist_Est}
\end{figure}

Figure~\ref{Fig:Dist_Est} illustrates the disturbance torque estimates generated by the Kalman filter for the three MPC test cases. 
The black dashed lines are the true disturbances, accounting for the torque generated by the SRP due to the non-ideal sail shape and the solar sail's attitude. The forward slew and return slew are initiated at $10000$ and $20000$ seconds respectively, which results in the change of SRP disturbance torque.
While the exact magnitude of the estimated disturbance torque does not exactly match the true disturbance torque, the estimate is reasonably accurate, and clearly assists with the MPC-based momentum management strategy, as shown in Fig.~\ref{Fig_MPC_sim}. 
It is worth noting that the disturbance torque estimate generated by the Kalman filter will account for all model inaccuracies in practice (e.g., nonlinearities, discretization approximations), which could explain the difference between the estimated and true disturbance torque. The disturbance torque in roll axis has a significantly smaller magnitude than the pitch/yaw axes and the other state estimates, making it difficult to observe and sensitive to measurement noise. However, the roll disturbance estimate still converges within the neighborhood of the true value, and allows MPC to adjust accordingly. Future work will investigate improving the observability of the roll disturbance estimate, through the use of more accurate measurements or the introduction of additional measurements.

Given the prediction model being linearized about the nominal slew trajectory, the MPC policies proactively take momentum management actuation once the slew maneuver arises in the prediction horizon (1000 seconds ahead in this case). 
Within the MPC formulation, the RCD torque magnitude, SRP force, estimated disturbance torque are assumed constant, with the AMT actuation modeled as a ZOH. In contrast, the simulation incorporates the nonlinear attitude dynamics, AMT motion dynamics, and the attitude-dependent nature of the SRP force, torque, and RCD effects. Despite the simplifications in the prediction model of MPC (which improves real-time feasibility), the recursive nature allows the controller to compensate for these discrepancies at every time step, demonstrating significant robustness against model uncertainties.

\subsection{State-of-the-Art Comparison} \label{sec:NumSim_Comparison}

A maneuver sequence tracking initial attitude of $\mbs{\theta}_0 = \mbf{0}$ to ${\mbs{\theta}}_\text{goal} = \Big[ 0^\circ \,\,\, 10^\circ \,\,\, 1^\circ \Big]^\trans$ following the desired hold-slew-hold-slew-hold trajectory is executed using the MPC framework to directly compare its actuation efficiency to that of NASA's state-of-the-art method~\citep{Inness2023MM,Tyler2024}.
This threshold-based control policy is a recreation of the momentum management logic for Solar Cruiser based on the information provided by~\cite{Inness2023MM,Tyler2024}. While it serves as a functional approximation for the purposes of this study, it is not an exact replica the proprietary controller implemented on NASA's flight hardware.
For this simulation, MPC uses the same tuning parameters as the simulations in Fig.~\ref{Fig_ThrMPC_sim} with $30\%$ AMT threshold and $90\%$ RCD threshold.

\begin{figure}[t!] 
\centering
\subfigure[attitude]
{
        \includegraphics[width=0.48\textwidth]{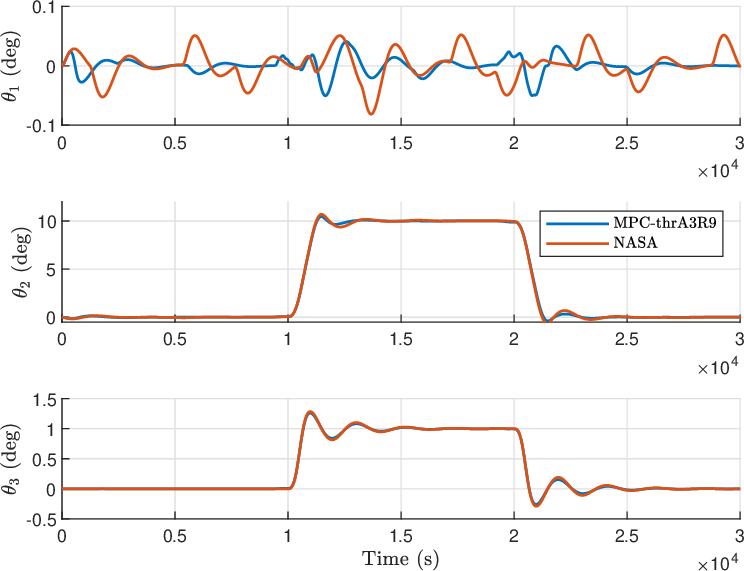}
        \label{Fig_NASAMPC_sima}
}
\subfigure[4 RWs angular momentum]
{
        \includegraphics[width=0.48\textwidth]{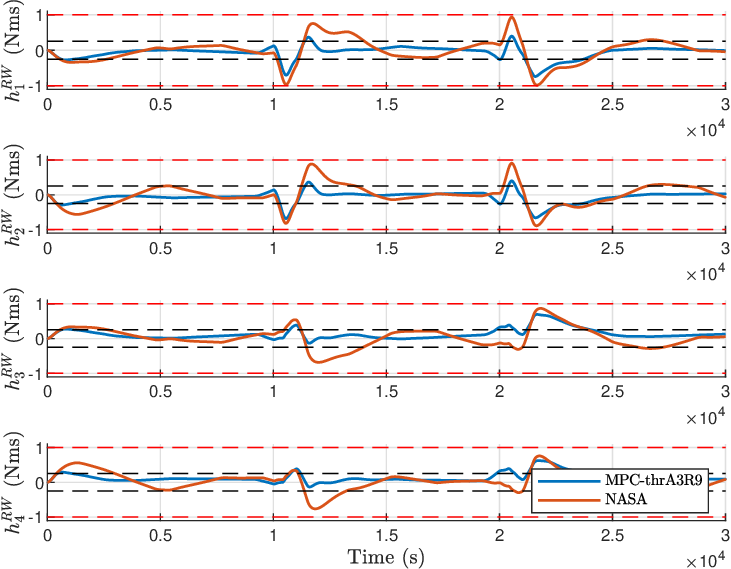}
        \label{Fig_NASAMPC_simb}
}
\\
\centering
\subfigure[body-frame RWs angular momentum]
{
        \includegraphics[width=0.48\textwidth]{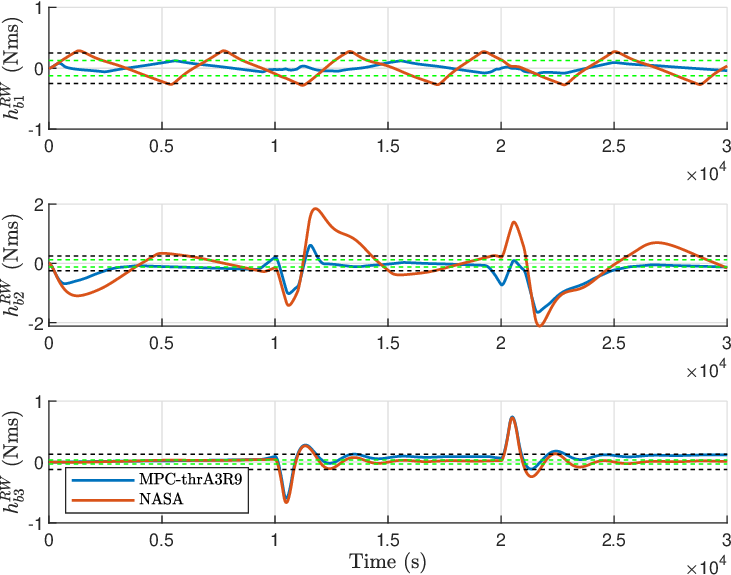}
        \label{Fig_NASAMPC_simc}
}
\subfigure[momentum management inputs]
{
        \includegraphics[width=0.48\textwidth]{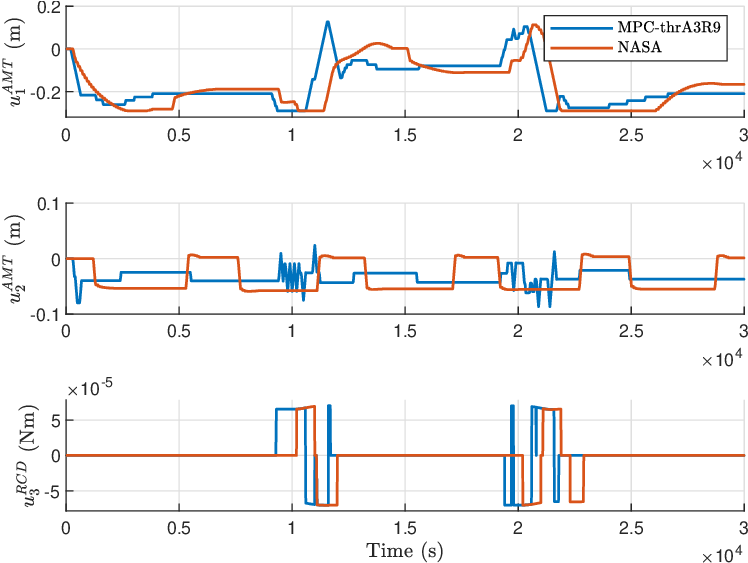}
        \label{Fig_NASAMPC_simd}
}
\centering
\caption{Comparison of simulation results using the proposed MPC momentum management strategy versus NASA's state-of-the-art method under a maneuver sequence tracking initial attitude $\mbs{\theta}_0 = \mbf{0}$ to ${\mbs{\theta}}_\text{goal} = \Big[ 0^\circ \,\,\, 10^\circ \,\,\, 1^\circ \Big]^\trans$ and return.}
\label{Fig_NASAMPC_sim}
\end{figure}

Figure~\ref{Fig_NASAMPC_sim} includes the comparison of simulation results using NASA's method~\citep{Inness2023MM,Tyler2024} and the proposed MPC approach with thresholds.
In Fig.~\ref{Fig_NASAMPC_simd}, the MPC proactively actuates the AMT and RCDs to avoid angular momentum growth, as shown in Figures~\ref{Fig_NASAMPC_simb} and~\ref{Fig_NASAMPC_simc}.
A quantitative comparison of the control actuation usage is included in Table~\ref{tab:NASAMPC_CtrlUsage}. 
The proposed MPC policy achieves a significant reduction in AMT and RCD usage during the attitude hold. In contrast, a higher actuation usage during the slew maneuver provides a significantly more effective momentum management and a broader operation region (i.e., a larger range of slew maneuvers in which momentum management can be effectively performed.
The PWM-quantization evenly distributes the input across every time step, as opposed to the longer singular ``on'' pulse with a long ``off'' period when using NASA's benchmark method.
Although the MPC results in higher RCD on-off cycles due to this inherent PWM quantization, dividing a long activation command into multiple short pulses is not inherently detrimental, as it mitigates the risk of potentially overheating the actuator associated with excessively long RCD ``on'' commands. Future work could investigate the design of an actuation mechanism capable of grouping these short MPC-generated pulses into a single, longer RCD activation event according to the mission requirements and hardware limitations.

\begin{table}[t!]
\caption{Control actuation usage with NASA Solar Cruiser's state-of-the-art momentum management method and the proposed MPC method with $30\%$ AMT and $90\%$ RCD threshold under a maneuver sequence from $\mbs{\theta}_0 = \mbf{0}$ to ${\mbs{\theta}}_\text{goal} = \Big[ 0^\circ \,\,\, 10^\circ \,\,\, 1^\circ \Big]^\trans$ and return.}
\label{tab:NASAMPC_CtrlUsage}
\centering
\begin{tabular}{c|cc|cc}
    Time & \multicolumn{2}{c|}{Slew Maneuver (0-30000 s)} & \multicolumn{2}{c}{Attitude Hold (30000-60000 s)} \\
    \cline{1-5}
    Controller & Solar Cruiser & MPC w/ Threshold & Solar Cruiser & MPC w/ Threshold \\
    \hline
    RCD Cycle ($\#$) & 5 & 10 & 1 & 0 \\
    RCD On Time (sec) & 3900 & 4695 & 1700 & 0 \\
    AMT Dist 1 (cm) & 171.48 & 208.75 & 63.26 & 9.53 \\
    AMT Dist 2 (cm) & 62.24 & 142.22 & 62.05 & 12.65 \\
    Sum of AMT Dist (cm) & 233.72 & 350.97 & 125.31 & 22.18 \\
\end{tabular}
\end{table}


\section{Conclusions}
\label{sec:Conclusions}

This paper presented a novel Kalman filter augmented MPC framework specifically designed for the challenging momentum management task of NASA's Solar Cruiser. The integrated estimation framework proposed in this work plays a crucial role, providing real-time state and disturbance estimates that not only characterize the external disturbance torque but also capture the dynamic discrepancies between the linear prediction model and the highly nonlinear spacecraft system. This estimate closes the modeling gap needed to enable model predictive control for this momentum management application. Building upon a previously-developed MPC architecture~\citep{shen2025}, the policy was rigorously formulated to be computationally feasible, utilizing off-the-shelf QP solvers to ensure real-time implementation capability within the limited onboard hardware. Validation with a more realistic SRP model, while performing slew maneuver following and incorporating Solar Cruiser's 4-RW configuration in this paper presents a key contribution towards the development of a practical implementation of the proposed MPC-based momentum management policy.

Simulation results demonstrated the proposed MPC-based momentum management policy's superior performance and robustness. The disturbance estimate was shown to be essential in achieving reliable MPC prediction and bounded momentum management. Furthermore, the proposed MPC policy successfully managed angular momentum growth under maneuvers that exceed the capability of NASA's state-of-the-art method designed for Solar Cruiser, establishing a larger operational slew envelope. It is shown that the MPC policy proactively unloads angular momentum in preparation for upcoming high-demand slew maneuvers. The framework also proved its efficiency by demonstrating reduced actuator usage through a lower AMT travel distance and optimized RCD usage compared to the benchmark method. This improvement has the potential to enable greater solar sail mission longevity.

Future work on this topic could be the investigation of improving roll disturbance estimates and reducing the number of RCD on-off cycles. Additional work towards the implementation of the proposed method on flight hardware and software will also be pursued to move towards its implementation on future solar sail missions.

\section*{Acknowledgments}
This material is based upon work supported by NASA under award No. 80NSSC25M7060, as well as a study grant from Chung Cheng Institute of Technology, National Defense University, Taiwan (R.O.C.). The authors would like to thank Mr. Keegan R. Bunker for the valuable discussions on sail shape modeling and providing the SRP model used in this work.

\section*{Appendix: Reference Trajectory} \label{sec:Appendix}
For the reference maneuver in this work, the design variables include the initial attitude $\mbs{\theta}_0$, the predetermined goal attitude $\mbs{\theta}_\text{goal}$, the maximum allowable slew rate $\dot{\mbs{\theta}}_\text{slew} \geq \mbf{0}$, and the acceleration time $t_\text{accel}$ required to reach the maximum slew rate. The slew maneuver is classified into acceleration, coast, and deceleration phases for both the forward maneuver start at $t_\text{start}$ and the return maneuver initiating at $t_\text{return}$. 

The reference trajectory holds at its initial state $\mbs{\theta}_d(t) = \mbs{\theta}_0$ and $\dot{\mbs{\theta}}_d(t) = \mbf{0}$ for $t<t_\text{start}$. 
The system then starts the forward maneuver, slewing toward $\mbs{\theta}_\text{goal}$, at $t_\text{start}$. 
The constant design variables $\dot{\mbs{\theta}}_\text{slew}$ and $t_\text{accel}$ define a constant angular acceleration magnitude $|\ddot{\theta}_{d,i}| = \dot{\theta}_{\text{slew},i} / t_{\text{accel}}$ for the $i$-th axis. 
During the acceleration phase where $t - t_\text{start} < t_{\text{accel}}$, the angular rate magnitude increases linearly following $|\dot{\theta}_{d,i}| = |\ddot{\theta}_{d,i}| (t - t_\text{start})$. 
Concurrently, the desired attitude evolves quadratically from the initial state as $\theta_{d,i} = \theta_{0,i} \pm \frac{1}{2} |\ddot{\theta}_{d,i}| (t - t_\text{start})^2$, where the $\pm$ sign depends on the slew direction defined by $\tilde{\theta}_{i} = \theta_{\text{goal},i} - \theta_{0,i}$.
Once the maximum slew rate is reached, the system enters a coast phase where $|\dot{\theta}_{d,i}| = \dot{\theta}_{\text{slew},i}$ is maintained. 
The constant angular velocity is held for $t_\text{coast} = 2\tilde{\theta}_{i}/\dot{\theta}_{\text{slew},i} -2t_\text{accel}$, until the remaining angular distance dictates the start of the deceleration phase. 
During the deceleration phase, a constant negative acceleration of magnitude $|\ddot{\theta}_{d,i}|$ is applied to smoothly bring the angular rate $\dot{\theta}_{d,i}$ back to zero exactly as the attitude reaches $\theta_{\text{goal},i}$.

The area under the angular velocity versus time plot denotes the total attitude angle change.
This integrated area matches the magnitude of slew angular displacement $|\tilde{\theta}_{i}|$. 
This geometric relationship establishes a strict kinematic constraint that dictates the angular velocity profile.
If the slew angle is large enough to complete the full acceleration and deceleration ramps, the standard trapezoidal profile is executed. 
If the required slew angle is too small, the maximum slew rate cannot be achieved without overshooting the target. Under this condition, the reference trajectory degenerates into a triangular velocity profile. The system accelerates and shortly begins deceleration before $t = t_\text{start} + t_\text{accel}$ without ever entering a constant velocity coast phase. 
To satisfy the exact slew angle constraint, the acceleration duration for the degenerated triangular velocity profile is calculated as $t_{\text{accel}\Delta} = \sqrt{|\tilde{\theta}_{i}| / |\ddot{\theta}_{d,i}|}$. The peak angular rate of the slew is thereby reduced to $|\ddot{\theta}_{d,i}| t_{\text{accel}\Delta}$. 

Following the completion of the forward maneuver, the system settles and holds at the goal state $\mbs{\theta}_d = \mbs{\theta}_\text{goal}$, $\mbs{\omega}_d = \mbf{0}$. 
This reference orientation is maintained until the specified return time $t_\text{return}$ triggers an identical but reversed kinematic sequence to drive the attitude from $\mbs{\theta}_\text{goal}$ back to the initial state $\mbs{\theta}_0$ for the interval $t_{\text{complete}} < t < t_{\text{return}}$, where $t_{\text{complete}} = t_{\text{start}} + |\tilde{\theta}_{i}| / \omega_{\text{slew},i} + t_{\text{accel}}$ when the slew angle is large enough to execute the standard trapezoidal profile, and $t_{\text{complete}} = t_{\text{start}} + 2 t_{\text{accel}\Delta}$ when the slew angle is insufficient and the trajectory degenerates into a triangular profile. This desired slew maneuver trajectory is formulated as a function that calculates the associated reference $\mbs{\theta}_d(t)$, $\dot{\mbs{\theta}}_d(t)$, and $\ddot{\mbs{\theta}}_d(t)$ at any given time $t$. In the RW PID control law, the desired angular momentum is calculated as $\mbs{\omega}_d(t) = \mbf{S}(\mbs{\theta}_d(t))\dot{\mbs{\theta}}_d(t)$, where the mapping matrix of a 3-2-1 Euler angle sequence is defined as 
\bdis
\mbf{S}(\mbs{\theta}) = \bbm 1 & 0 & -\sin(\theta_2) \\ 0 & \cos(\theta_1) & \sin(\theta_1)\cos(\theta_2) \\ 0 & -\sin(\theta_1) & \cos(\theta_1)\cos(\theta_2) \ebm.
\edis

\bibliographystyle{elsarticle-harv}
\biboptions{authoryear}
\bibliography{isss.bib}

\end{document}